\documentclass[11pt]{article}
\usepackage[utf8]{inputenc}

\usepackage{amsthm, amsfonts, amssymb, amsmath,enumitem, natbib, bbm, mathabx}
\usepackage{wrapfig}
\usepackage{cases, amsthm}
\usepackage{fullpage}
\usepackage{mathtools}
\mathtoolsset{showonlyrefs}
\usepackage{ wasysym} 
\usepackage{fancyhdr}
\usepackage{graphicx}
\usepackage{bbm}
\usepackage{caption}
\usepackage{subcaption}
\usepackage{algorithm}
\usepackage{algorithmic}
\usepackage{authblk}
 \usepackage{setspace}
\let\Algorithm\algorithm
\renewcommand\algorithm[1][]{\Algorithm[#1]\setstretch{1.6}}

\usepackage[colorlinks=true, pdfstartview=FitV,
linkcolor=blue, citecolor=blue, urlcolor=blue]{hyperref}

\newtheorem{definition}{Definition}[section]
\newtheorem{theorem}[definition]{Theorem}

\newtheorem{exmp}{Example}[section]

\newcommand{\N}{\mathbbm{N}}

\newcommand{\re}{\mathbb{R}}

\newcommand{\indep}{\perp \!\!\! \perp}

\DeclareMathOperator*{\ord}{\mathcal{O}}
\newcommand{\al}{\alpha}
\newcommand{\del}{\delta}
\newcommand{\Ical}{\mathcal{I}_{\mathrm{cal}}}

\usepackage{xcolor}
\DeclareMathOperator{\one}{\mathbf{1}} 

\newcommand{\cov}{\mathsf{Cov}}
\newcommand{\prob}{\mathbb{P}}

\renewcommand{\algorithmicrequire}{\textbf{Input:}}

\newcommand{\eps}{\epsilon}

\newcommand{\wh}{\widehat}  

\def\R{\mathbb{R}}
\def\Pr{\mathbb{P}}

\def\R{\mathbb{R}}
\def\Pr{\mathbb{P}}

\usepackage[top=1in, bottom=1in, left=1in, right=1in]{geometry}
\title{Post-selection Inference for Conformal Prediction:\\Trading off Coverage for Precision}

\author[1]{Siddhaarth Sarkar}
\author[1]{Arun Kumar Kuchibhotla}
\affil[1]{Department of Statistics \& Data Science, Carnegie Mellon University}
\date{}

\begin{document}

\maketitle

\begin{abstract}
Conformal inference has played a pivotal role in providing uncertainty quantification for black-box ML prediction algorithms with finite sample guarantees. Traditionally, conformal prediction inference requires a data-independent specification of miscoverage level. In practical applications, one might want to update the miscoverage level after computing the prediction set. For example, in the context of binary classification, the analyst might start with a 95$\%$ prediction sets and see that most prediction sets contain all outcome classes. Prediction sets with both classes being undesirable, the analyst might desire to consider, say 80$\%$ prediction set. Construction of prediction sets that guarantee coverage with data-\emph{dependent} miscoverage level can be considered as a post-selection inference problem. In this work, we develop simultaneous conformal inference to account for data-dependent miscoverage levels. Under the assumption of independent and identically distributed observations, our proposed methods have a finite sample simultaneous guarantee over all miscoverage levels. This allows practitioners to trade freely coverage probability for the quality of the prediction set by any criterion of their choice (say size of prediction set) while maintaining the finite sample guarantees similar to traditional conformal inference.   
\end{abstract}

\section{Introduction}

\subsection{Conformal inference}
Machine learning algorithms are focused on achieving 
the best possible point prediction for a given dataset containing a set of covariates $X$ and response $Y$. One approach is to make assumptions about the joint distribution of $(X,Y)$ and mimic the optimal regression or classification function; this includes classical parametric prediction algorithms such as linear/logistic regression and also some non-parametric prediction algorithms such as generalized additive models (GAMs). But these methods heavily rely on distributional assumptions for predictive inference. Another approach is to develop methods that yield predictive inference robust to the modeling assumptions. Thus, practical procedures for providing uncertainty quantification for black-box methods without any distributional assumptions are important. Conformal  prediction methods \citep{vovk1999machine,lei2013distribution} achieve this with non-asymptotic coverage guarantees that are agnostic of the distribution of data and the prediction algorithm. Given data $\mathcal{D}_n = \{Z_i: i \in [n]\}$ with independent and identically distributed random vectors from $P$ and a desired coverage rate $1 - \al \in (0, 1)$, conformal prediction methods construct a set $\wh{C}_{n,\al}$ from $\mathcal{D}_n$ such that for an independent $Z_{n+1}\sim P,$
$\mathbb{P}(Z_{n+1} \in \wh{C}_{n,\al}) \geq 1- \al.$
This is called marginal coverage guarantee.
Here the probability is computed with respect to $Z_{n+1}$ and $\widehat{C}_{n,\alpha}$.

Arguably, one of the most popular conformal prediction methods is the split conformal method which is based on sample splitting and one-dimensional projection (called the non-conformity score). The split conformal method is central to our proposed methodology and hence we briefly discuss the idea now. Suppose the given data $\mathcal{D}_n = \{Z_i:\,1\le i\le n\}$ consisting of IID random variables from a measurable space $\mathcal{Z}$ be split into two independent parts: training data $\mathcal{D}_{\mathrm{tr}} = \{Z_i:\,i\in\mathcal{I}_{\mathrm{tr}}\}$ (for some $\mathcal{I}_{\mathrm{tr}} \subseteq\{1, 2, \ldots, n\}$) and calibration data $\mathcal{D}_{\mathrm{cal}} = \{Z_i:\, i\in\mathcal{I}_{\mathrm{cal}} = \mathcal{I}_{\mathrm{tr}}^c\}$. Let $s:\mathcal{Z}\to\mathbb{R}$ be any function (called non-conformity score or conformal score) that maps $\mathcal{Z}$ to $\mathbb{R}$ and is obtained only based on $\mathcal{D}_{\mathrm{tr}}$. For example, one can take $s(z) = 1/\widehat{p}_{\mathrm{tr}}(z)$ where $\widehat{p}_{\mathrm{tr}}(\cdot)$ is some estimator of the density of $Z_i$'s~\citep{kim2019uniform}. Note that, in this example, $s(z)$ measures non-conformity of the point $z$ to the training data $\mathcal{D}_{\mathrm{tr}}$ because high-density points correspond to the majority of the training data $\mathcal{D}_{\mathrm{tr}}$ and low-density points usually correspond to those that are ``far'' from $\mathcal{D}_{\mathrm{tr}}$. In the next step of the split conformal method, one defines 
$\widehat{Q}_{n,\alpha}$ as the $\lceil (|\mathcal{I}_{\mathrm{cal}}| + 1)(1-\alpha)\rceil$-th largest value of $s(Z_i), i\in\mathcal{I}_{\mathrm{cal}}.$ 
($\widehat{Q}_{n,\alpha}$ is taken to be infinity if $(|\mathcal{I}_{\mathrm{cal}}| + 1)(1-\alpha) > |\mathcal{I}_{\mathrm{cal}}|$.)
The split conformal prediction set is given by $\widehat{C}_{n,\alpha} = \{z:\, s(z) \le \widehat{Q}_{n,\alpha}\}.$ No matter what the underlying space $\mathcal{Z}$, the distribution of $Z_i$'s, and score $s(\cdot)$ are, the prediction set $\widehat{C}_{n,\alpha}$ satisfy marginal coverage guarantee~\citep{lei2013distribution,romano2019conformalized}. Although the intuition of $s(\cdot)$ as a measure of non-conformity is useful in deriving a ``good'' prediction set, this is not needed for the validity guarantee of $\widehat{C}_{n,\alpha}$. 
The marginal coverage guarantee should be contrasted with the conditional coverage guarantee: $\mathbb{P}(Y_{n+1} \in \wh{C}_{n,\al}(X_{n+1}) \big|  X_{n+1} = x) \geq 1- \al,$ for any $x$, which is known to be impossible to attain without strong assumptions on the joint distribution \citep{foygel2021limits}. Several authors \citep{sesia2020comparison, kivaranovic2020adaptive, romano2019conformalized, romano2020classification} constructed procedures that attain finite sample marginal coverage guarantee but also asymptotic conditional coverage guarantee using conditional quantile regression or conditional probability estimators. 

Of course, in considering the usage of a particular conformal prediction method, one should also account for the width of the prediction sets thus obtained. The choice of the non-conformity score used in the split conformal method plays a very crucial role in width considerations. \cite{lei2013distribution, lei2018distribution,sesia2020comparison}
show that conformal prediction sets constructed from particular conformal scores can be asymptotically optimal in terms of width/Lebesgue measure. In an unsupervised setting,~\cite{lei2013distribution} proved that inverse density score yields a finite-sample valid prediction set that converges to the optimal prediction set assuming that the density estimator is consistent. Similarly, in a supervised regression setting, under a strong assumption on the error distribution, \cite{lei2018distribution} showed that using $s(z) = |y - \widehat{\mu}(x)|$ (for $z = (x, y)$) can yield the optimal prediction set. \cite{sesia2020comparison} relaxed these strong assumptions and proved that non-conformity scores constructed using conditional quantile regression estimators can yield prediction sets that are optimal.

Conformal prediction in recent years has been extended to various settings including covariate shift or missing data \citep{tibshirani2019conformal,lei2021conformal, yang2022doubly, barber2022conformal}, censored data \citep{candes2023conformalized, gui2022conformalized}, continuous distributional shifts \citep{gibbs2021adaptive,gibbs2022conformal,barber2022conformal,bhatnagar2023improved}, multilabel data \citep{cauchois2020knowing}, network-based covariates \citep{lunde2023conformal}, and time-series \citep{zaffran2022adaptive, xu2021conformal}. All of these methods yield prediction sets that contain the true response with a \textit{nominal level of confidence} either for all sample sizes or asymptotically. All these papers use marginal coverage guarantee with a fixed level of miscoverage error $\alpha.$ In the conformal prediction literature, other criteria of coverage guarantee are also considered. A brief discussion of some of these alternative criteria is useful for understanding our goal.
\subsection{Different types of guarantees.}
Marginal coverage guarantee can be understood from a decision theory perspective as the expected coverage probability. 
We can formally define it as follows. 
\begin{definition}\label{def:cov_prob}\textbf{[Coverage probability]} Given a prediction set $\wh{C}_{n,\al}$, the coverage probability attained for an independent $Z\sim P$ is defined as $\cov_P(\wh{C}_{n,\al}) = \prob_{Z\sim P}(Z \in \wh{C}_{n,\al}\mid\wh{C}_{n,\al}).$
\end{definition}

Note that $\cov_P(\wh{C}_{n,\al})$ is a random quantity. 
The marginal coverage guarantee is now the same as $\mathbb{E}[ \cov_P(\wh{C}_{n,\al}) ] \geq 1- \al$. Instead of ensuring a lower bound on the expectation of $\cov_P(\wh{C}_{n,\al})$, one could ask for a high probability lower bound of $1 - \al$ for $\cov_P(\wh{C}_{n,\al})$. Given a desired high probability level, say $ 1- \delta$ with $\delta \in (0,1)$, one could require
 \begin{equation}\label{eq:pac-guarantee}
     \Pr(\cov_P(\wh{C}_{n,\al}^\del) \geq 1- \al) \geq 1 - \delta.
 \end{equation}
 This is referred to as $(\al,\delta)$-probably approximately correct (PAC) guarantee and is discussed in \cite{valiant1984theory,guttman1967statistical,krishnamoorthy2009statistical}. The idea behind $(\alpha, \delta)$-PAC is to roughly ensure that on replicating the experiment several times, $100(1-\delta)\%$ of the experiments will have coverage higher than $1-\al$. 
\subsection{Limitations of the conformal framework.}
Conformal methods are reliant on a fixed choice of coverage probability $1-\al$. This choice has to be made prior to obtaining the prediction region and independently of the data. But if this coverage probability itself is data-dependent, then the coverage guarantees are no longer valid. This problem is called \textit{selection bias} and often discussed in statistical settings, such as hypothesis testing \citep{olshen1973conditional}, PCA \citep{choi2017selecting}, change point detection \citep{hyun2016exact} among others. In the standard framework, if the practitioner wants to change the coverage level \textit{after} seeing that the procedure gives an undesirable amount of precision, they would lose validity. There is a trade-off between precision (width/size of prediction set) and coverage level. This trade-off was suggested as a useful practical tool in~\cite{kuchibhotla2023nested}. Not having a framework for a data-dependent $\al$ can lead to erroneous results if a practitioner chooses $\al$ in a data-dependent manner. It is worth highlighting that conformal prediction in the context of selection has been considered in the literature, although the selection is mostly related to test subjects/units; see, for example,~\cite{bates2023testing} and~\cite{jin2022selection}.



In the following sections, we propose to construct prediction sets with data-dependent miscoverage levels using simultaneously valid confidence intervals for a cumulative distribution function. This can be considered analogous to the simultaneous inference framework of post-selection inference~\citep{berk2013valid}. An alternative to our proposed solution can be obtained analogously to sample splitting. We will only briefly describe this sample splitting method here and not explore this further as it requires further splitting of data. Divide the dataset into three parts: training, calibration, and precision set. Then, an $\al = \widehat{\alpha}$ is picked based on any independent precision metric (such as average width/size of prediction set on the precision set) on prediction sets computed only using the training and calibration sets. 
This method leads to a smaller dataset to train and build the conformal set. Moreover, while the conformal set is valid, it may not provide the exact level of precision used to select $\alpha$ when applied to new test data points. Referring to the earlier example, say the practitioner picks an $\al$ such the average size of the prediction set is $1$ on the precision set. This will not ensure that we get a prediction set of size $1$ on a new test datapoint for that $\al$, albeit it is statistically close to $1$. 

\paragraph{Organization.} The remaining article is organized as follows. In Section~\ref{sec:prob_formulation}, we formulate the post-selection inference problem for conformal prediction and using the split conformal prediction framework connect this to the problem of confidence bands for cumulative distribution functions. In Section~\ref{sec:methods}, we propose two methods to solve the post-selection inference problem and compare them. In Section~\ref{sec:simulations-real-data}, we study the performance of the proposed methods on simulated and real datasets. Finally, we conclude the article with a summary and a discussion of some future directions in Section~\ref{sec:conclusion}.
\section{Problem Formulation}\label{sec:prob_formulation}
\subsection{Simultaneous PAC guarantee.}
Selective inference procedures \citep{berk2013valid} try to address questions of data dependence choices by providing guarantees for all possible choices that one can make. In a similar spirit, one could ask for a similar guarantee in the PAC setting as follows,
 \begin{equation}\label{eq:simul-pac-guarantee}
   \Pr\left(\cov_P(\wh{C}_{n,\al}^\del) \geq 1- \al \quad \forall \ \al \in (0,1)\right) \geq 1 - \delta.
 \end{equation}
We refer to this as \textit{simultaneous $\delta$-PAC guarantee}. 
This guarantee would automatically give us that if any $\wh{\al}$ is chosen based on $\mathcal{D}_n$, then we can say that $\Pr(\cov_P(\wh{C}_{n,\wh{\al}}^\del) \geq 1- \wh{\al}) \geq 1 - \delta. $
A simultaneous PAC guarantee would allow practitioners  to trade coverage for precision in a valid manner. Moreover, if the analyst does not require a guarantee for all $\alpha\in(0, 1)$, but only for a target set $I \subseteq (0,1)$ of miscoverage levels, then we consider simultaneous $(I, \delta)$-PAC guarantee: 
    $\Pr(\cov_P(\wh{C}_{n,\al}^\del) \geq 1- \al \quad \forall \ \al \in I) \geq 1 - \delta.$
For example, if the practitioner only wants to trade off miscoverage levels up to $0.5$, then the simultaneous ([0, 0.5],$\delta$)-guarantee is required. Setting $I=\{\alpha\}$ gives the usual $(\alpha,\delta)$-PAC guarantee discussed in~\cite{vovk2012conditional,bian2022training}.

Using prediction sets constructed through non-conformity scores, one can convert prediction sets that satisfy simultaneous $\delta$-PAC guarantee using distribution-free confidence bands for univariate cumulative distribution functions.
\subsection{CDF Confidence bands and simultaneous guarantee.}
A random variable $X \in \R$ has a CDF $F: \mathbbm{R} \to \mathbbm{R}$ which is defined as $F(x) = \prob(X\leq x)$. Given an IID sample, $X_1, \ldots, X_n$ from this distribution $F$ and $\del \in (0,1)$, a \textit{CDF confidence band} construction refers to constructing random functions $\ell_{n,\delta}(x), u_{n,\delta}(x)$ such that 
$$ \Pr( \ell_{n,\delta}(x) \leq F(x) \leq u_{n,\delta}(x) \quad \forall\ x) \geq 1 - \del .$$
Note that a ``good'' CDF confidence band should satisfy $\ell_{n,\delta}(x) \to 0$ as $x\to -\infty$ and $u_{n,\delta}(x) \to 1$ as $x \to \infty$. The construction of a valid $1-\delta$ CDF confidence band implies prediction sets that satisfy simultaneous $\delta$-PAC guarantee. This can be achieved as follows, using the non-conformity score. 

Recall that in the split conformal method, one part of the data is used to construct a conformal score $s(\cdot)$, and the other part yields scores $s(Z_i), i\in\mathcal{I}_{\mathrm{cal}}$. Also, recall that the resulting conformal prediction set is given by $\widehat{C}_{n,\alpha} := \{z\in\mathcal{Z}:\, s(z) \le \widehat{Q}_{n,\alpha}\}.$
In order to obtain prediction sets that satisfy simultaneous $\delta$-PAC guarantee, we change $\widehat{Q}_{n,\alpha}$ using the confidence band. Consider
$\widehat{C}_n(t) = \{z\in\mathcal{Z}:\, s(z) \le t\}.$
It is clear that for $Z\sim P$, $\cov_P(\widehat{C}_{n}(t)) = \mathbb{P}(s(Z) \le t),$ and hence, the coverage probability is the value of the CDF of $s(Z)$ at $t$. If $[\ell_{n,\delta}(t), u_{n,\delta}(t)], t\in\mathbb{R}$ is a confidence band for $t\mapsto \mathbb{P}(s(Z) \le t)$, based on the scores $s(Z_i), i\in\mathcal{I}_{\mathrm{cal}}$, i.e., $\ell_{n,\delta}(t) ~\le~ \cov_P(\widehat{C}_n(t)) ~\le~ u_{n,\delta}(t)$ holds simultaneously for all $t\in\mathbb{R}$ with probability at least $1 - \delta$, then one can obtain prediction sets with simultaneous $\delta$-PAC guarantee as shown in Algorithm~\ref{algo:simul-pac}. Theorem~\ref{thm:si-PAC-guarantee} proves that the confidence sets returned by Algorithm~\ref{algo:simul-pac} satisfy the simultaneous $\delta$-PAC guarantee.
\begin{algorithm}[t]
\caption{Simultaneous $\delta$-PAC conformal prediction}
\label{algo:simul-pac}
\begin{algorithmic}[1]

\REQUIRE{$\delta \in (0,1)$, and data $Z_1, \ldots, Z_n\in\mathcal{Z}$.
 , and a confidence band $[\ell_{n,\delta}(t), u_{n,\delta}(t)], t\in\mathbb{R}$ for the CDF of $s(Z)$.
}
\renewcommand{\algorithmicrequire}
{\textbf{Procedure:}}
\REQUIRE{}
\STATE Split the data $Z_i, i\in[n]$ into (a) training data $Z_i, i\in\mathcal{I}_{\mathrm{tr}}$ and (b) calibration data $Z_i,\in\mathcal{I}_{\mathrm{cal}}$.
\STATE Construct the non-conformity score $s:\mathcal{Z}\to\mathbb{R}$ based on training data $Z_i, i\in\mathcal{I}_{\mathrm{tr}}$.
\STATE Find non-conformity score $S_i = s(Z_i), i\in\mathcal{I}_{\mathrm{cal}}$ and find a confidence band $[\ell_{n,\delta}(t), u_{n,\delta}(t)], t\in\mathbb{R}$ for the CDF $F(t) = \mathbb{P}(s(Z) \le t|Z_i, i\in\mathcal{I}_{\mathrm{tr}})$.
\STATE For each $\alpha\in(0, 1)$, set $\widehat{Q}_{n,\alpha}^{\mathrm{sim}}$ such that $\ell_{n,\delta}(\widehat{Q}_{n,\alpha}^{\mathrm{sim}}) \ge 1 - \alpha$.
\STATE Define prediction sets $\widehat{C}_{n,\alpha}^{\text{sim}} := \widehat{C}_{n}(\widehat{Q}_{n,\alpha}^{\mathrm{sim}}) = \{z\in\mathcal{Z}:\, s(z) \le \widehat{Q}_{n,\alpha}^{\mathrm{sim}}\}.$
\ENSURE{Return the simultaneous $\delta$-PAC prediction sets $\{\widehat{C}_{n,\alpha}^{\text{sim}}\}_{\alpha\in(0, 1)}$.}
\end{algorithmic}
\end{algorithm}
\begin{theorem}\label{thm:si-PAC-guarantee}
Suppose $Z_i, i\in\mathcal{I}_{\mathrm{cal}}, Z$ are identically distributed random variables in some measurable space $\mathcal{Z}$. Set $F(t) = \mathbb{P}(s(Z) \le t|Z_i, i\in\mathcal{I}_{\mathrm{tr}})$. If $\mathbb{P}(\ell_{n,\delta}(t) \le F(t) \le u_{n,\delta}(t)\;\forall\;t\in\mathbb{R}) \ge 1 - \delta$, then for $\{\widehat{C}_{n,\alpha}^{\mathrm{sim}}\}_{\al\in(0,1)}$ constructed by Algorithm~\ref{algo:simul-pac}, 
\[
\mathbb{P}\left(1 - \alpha \le \cov_P(\widehat{C}_{n,\alpha}^{\mathrm{sim}}) \le 1 - \alpha + R_{n,\alpha}\quad\mbox{for all}\quad \alpha\in(0, 1)\right) \ge 1 - \delta,
\]
where $R_{n,\alpha} := [\ell_{n,\delta}(\widehat{Q}_{n,\alpha}^{\mathrm{sim}}) - (1 - \alpha)] + [u_{n,\delta}(\widehat{Q}_{n,\alpha}^{\mathrm{sim}}) - \ell_{n,\delta}(\widehat{Q}_{n,\alpha}^{\mathrm{sim}})] \ge 0$.
\end{theorem}


In the next section, we will discuss the construction of confidence bands for IID data and study how $R_{n,\alpha}$ behaves for these methods. The decomposition of $R_{n,\alpha}$ is very instructive. The first term of $R_{n,\alpha}$ shows how conservatively valid the resulting prediction set is, and the second term is a bound on conservativeness. 
 Before that, let us consider examples of how the procedure would change from the standard fixed $\al$ framework.
\begin{exmp}\label{eg:density} \textbf{Density Level Sets.}
Suppose $Z_1, \ldots, Z_n$ are independent and identically distributed random variables from some measurable space $\mathcal{Z}$. Construction of optimal prediction sets for a future random variable is considered in~\cite{lei2013distribution}. Split the data into two parts: $\mathcal{D}_{\mathrm{tr}}:= \{Z_i:\,i\in\mathcal{I}_{\mathrm{tr}}\}$ and $\mathcal{D}_{\mathrm{cal}}:= \{Z_i:\,i\in\mathcal{I}_{\mathrm{cal}}\}$. Construct a density estimator $\widehat{p}(\cdot)$ based on $\mathcal{D}_{\mathrm{tr}}$; this can be a kernel density estimator, for example~\citep{kim2019uniform}. Define
\[
\widehat{F}_n(t) = \frac{1}{|\mathcal{I}_{\mathrm{cal}}|}\sum_{i\in\mathcal{I}_{\mathrm{cal}}} \mathbf{1}\{s(Z_i) \le t\},\quad\mbox{where}\quad s(z) = \frac{1}{\widehat{p}(z)}.
\]
Also, recall $F(\cdot)$ from Theorem~\ref{thm:si-PAC-guarantee}. The traditional split conformal prediction set is given by
\[
\widehat{C}_{n,\alpha}^{\mathrm{split}} := \{z:\, s(z) \le \widehat{Q}_{n,\alpha}^{\mathrm{split}}\}\quad\mbox{where}\quad \widehat{Q}_{n,\alpha}^{\mathrm{split}}\mbox{ satisfies }\widehat{F}_n(\widehat{Q}_{n,\alpha}^{\mathrm{split}}) \ge \frac{\lceil(|\mathcal{I}_{\mathrm{cal}}| + 1)(1 - \alpha)\rceil}{|\mathcal{I}_{\mathrm{cal}}|}.
\]
The $(\alpha, \delta)$-PAC prediction set from Proposition 2a of~\cite{vovk2012conditional} is given by
\[
\widehat{C}_{n,\alpha}^{\mathrm{PAC}} := \{z:\, s(z) \le \widehat{Q}_{n,\alpha}^{\mathrm{PAC}}\}\quad\mbox{where}\quad \widehat{Q}_{n,\alpha}^{\mathrm{PAC}}\mbox{ satisfies }\widehat{F}_{n}(\widehat{Q}_{n,\alpha}^{\mathrm{PAC}}) \ge \frac{\lceil(|\mathcal{I}_{\mathrm{cal}}| + 1)(1 - \alpha)\rceil}{|\mathcal{I}_{\mathrm{cal}}|} + \sqrt{\frac{\log(1/\delta)}{2|\mathcal{I}_{\mathrm{cal}}|}}.
\]
This can be improved using the binomial distribution function as suggested in Proposition 2b of~\cite{vovk2012conditional}. 
The simultaneous $\delta$-PAC prediction set based on a lower confidence band $\ell_{n,\delta}(\cdot)$ is given by
\[
\widehat{C}_{n,\alpha}^{\mathrm{sim}} := \{z:\,s(z) \le \widehat{Q}_{n,\alpha}^{\mathrm{sim}}\}\quad\mbox{where}\quad\widehat{Q}_{n,\alpha}^{\mathrm{sim}}\mbox{ satisfies }\ell_{n,\delta}(\widehat{Q}_{n,\alpha}^{\mathrm{sim}}) \ge 1 - \alpha.
\]
A simple example of a lower confidence band can be obtained using the DKW inequality. For example, Comment 1 of~\cite{massart1990tight} implies that
\[
\mathbb{P}\left(|\mathcal{I}_{\mathrm{cal}}|^{1/2}\sup_{t}\{F(t) - \widehat{F}_n(t)\} \ge \lambda\right) \le \exp(-2\lambda^2)\quad\mbox{for all}\quad \lambda \ge \sqrt{\log(2)/2}.
\]
This, in particular, yields $\ell_{n,\delta}(t) = \widehat{F}_n(t) - \sqrt{\log(2/\delta)/(2|\mathcal{I}_{\mathrm{cal}}|)}$ as a valid lower confidence band for all $\delta\in(0, 1/2)$. In this example, with the simple PAC and simultaneous prediction sets have similar thresholds $\widehat{Q}_{n,\alpha}^{\mathrm{PAC}}$ and $\widehat{Q}_{n,\alpha}^{\mathrm{sim}}$, both of which are larger than $\widehat{Q}_{n,\alpha}^{\mathrm{split}}$. Note that all the thresholds converge to population quantile $F^{-1}(1 - \alpha)$ as $|\mathcal{I}_{\mathrm{cal}}| \to \infty$.
\end{exmp}
\begin{exmp} \label{eg:dcp}\textbf{Conformal prediction for regression.} 
Suppose $Z_i = (X_i, Y_i), 1\le i\le n$ are independent and identically distributed observations, with $Y_i$ being continuous. We now consider prediction sets that focus on predicting response with the side information of covariates.
In this setting, several authors have considered non-conformity scores that yield approximate conditional coverage, as discussed in the introduction. For this example, we use the distributional conformal score of~\cite{chernozhukov2021distributional}. As before, split data into two parts $\mathcal{D}_{\mathrm{tr}} = \{Z_i:\,i\in\mathcal{I}_{\mathrm{tr}}\}$ and $\mathcal{D}_{\mathrm{cal}} = \{Z_i:\,i\in\mathcal{I}_{\mathrm{cal}}\}$. Based on $\mathcal{D}_{\mathrm{tr}}$, estimate the conditional distribution function $H(y|x) = \prob( Y\leq y| X = x)$ with $\wh{H}(y|x)$. The non-conformity score is now given by $s(z) = |\wh{H}(y|x) - 1/2|$ for $z = (x, y)$. With this score, different types of prediction sets can be constructed, as in Example~\ref{eg:density}. Note that for $t\in[0, 1/2]$,
$\{z:\,s(z) \le t\} = \{(x, y):\,1/2 - t \le \wh{H}(y|x) \le 1/2 + t\}.$
\end{exmp}

\begin{exmp} \label{eg:class}\textbf{Conformal prediction for classification.} 
Suppose $Z_i = (X_i, Y_i), 1\le i\le n$ are independent and identically distributed observations, with $Y_i$ being categorical (for simplicity, we assume $Y_i \in \{1,\ldots , C\}$) . Similar to Example~\ref{eg:dcp}, among several methods for obtaining non-conformity scores, we choose \cite{romano2020classification}. Each datapoint is augmented with an independently generated (and thus always accessible) $U_i\sim\mathrm{Unif}[0,1]$ to allow for randomization. As before, split data into two parts $\mathcal{D}_{\mathrm{tr}} = \{Z_i:\,i\in\mathcal{I}_{\mathrm{tr}}\}$ and $\mathcal{D}_{\mathrm{cal}} = \{Z_i:\,i\in\mathcal{I}_{\mathrm{cal}}\}$. Based on $\mathcal{D}_{\mathrm{tr}}$, estimate the conditional probability mass function $\prob( Y = y| X = x)$ with $\wh{\pi}(y|x)$. Using $\wh{
\pi}(y|x)$, construct a generalized conditional quantile function
$$L(x; \pi, \tau ) = \min\{c \in \{1,\ldots , C\} : \pi_{(1)}(x) + \pi_{(2)}(x) + \ldots + \pi_{(c)}(x) \geq \tau \}$$
where $\pi_{(i)}(x)$ is the $i^{\text{th}}$ largest element in $\{\pi(y|x) : y \in  \{1,\ldots , C\}\}$. To extract the most likely set of labels using $L$, $S(x, u; \pi, \tau)$ is defined as
$$
S(x, u; \pi, \tau ) = \begin{cases}
\text{‘$y$’ indices of the $(L(x; \pi, \tau ) - 1)$ largest $\pi(y|x)$}, & u \leq  V (x; \pi, \tau ),\\
\text{‘$y$’ indices of the $L(x; \pi, \tau )$ largest $\pi(y|x)$}, & \text{otherwise},
\end{cases}
$$
where
$$   V (x; \pi, \tau )= \frac{\left( \sum_{c = 1}^{L(x;\pi,\tau)}\pi_{(c)}(x)\right) - \tau}{\pi_{(L(x;\pi,\tau))}(x)} $$
The $u\in[0,1]$ allows for randomization. With a $U\sim\mathrm{Unif}[0,1], S(x, U; \pi, \tau )$ will return  a random set of the most likely labels with a net probability (with respect to $\pi(\cdot|x)$) of exactly $\tau$. The non-conformity score for $(X,Y) \indep U\sim\mathrm{Unif}[0,1]$ is defined as
$$E(X, Y, U; \wh{\pi}) = \min \left\{\tau \in [0, 1] : Y \in S(X, U; \wh{\pi}, \tau )\right\}.$$
With this score, different types of prediction sets can be constructed as in Example~\ref{eg:density}. Note that for $t\in[0, 1]$, we get the following prediction set
$\{z:\,s(z) \le t\} = \{(x, y,u):\, E(x,y,u; \wh{\pi}) \le  t\} = \{(x, y,u):\, y \in S(x,u; \wh{\pi}, t) \}.$
\end{exmp}

\section{Methods}\label{sec:methods}
In this section, we describe more powerful finite-sample confidence bands for CDF given a sequence of independent random variables than the DKW confidence band. From the equivalence discussed above between confidence bands and simultaneous $\delta$-PAC guarantee, a sharper confidence band implies sharper simultaneous $\delta$-PAC prediction sets.

In the following subsections, we review and discuss three confidence bands for CDF of univariate IID random variables. To avoid notational conflict, we use $W_1, W_2, \ldots, W_m$ to denote $m$ IID random variables from a CDF $F_W(\cdot)$. In the context of conformal prediction, $W_i = s(Z_i), i\in\mathcal{I}_{\mathrm{cal}}$.
\subsection{Confidence bands: literature review}\label{subsec:conf_band_litrev}
A standard route to construct confidence bands is to use the empirical CDF. Given the IID sample $W_1, \ldots, W_m$ from a distribution $F_W$, the empirical CDF $\wh{F}_m$ is given by $\wh{F}_m(w) = m^{-1}\sum_{i=1}^m \mathbf{1}\{W_i \leq w\} \forall w \in \re,$
where $\one\{\cdot\}$ is the indicator function. The classic example of this is the DKW inequality, \citep{dvoretzky1956asymptotic,massart1990tight} 
which
implies that $[\ell_{m,\delta}(w), u_{m,\delta}(w)], w\in\mathbb{R}$ with $\ell_{m,\delta}(w) = \wh{F}_m(w) - \sqrt{\ln(2/\delta)/(2m)}$ and $u_{m,\delta}(w) = \wh{F}_m(w) + \sqrt{\ln(2/\delta)/(2m)}$ is a valid confidence band for $F_W(w), w\in\mathbb{R}$.
It is worth noting that this is a fixed-width confidence band, i.e., $u_{m,\delta}(w) - \ell_{m,\delta}(w) = \sqrt{2\ln(2/\delta)/m}$ does not depend on $w$. 
However, considering the asymptotics of $\widehat{F}_m(w)$, it is easy to see that a variable width confidence band can yield a higher power. For example, the central limit theorem implies that $\sqrt{m}(\wh{F}_m(w) - F_W(w)) \overset{d}{\to} N\big(0, F_W(w)(1 - F_W(w)) \big)$ as $m\to\infty$ and thus one can expect width to scale like $\sqrt{F_W(w)(1 - F_W(w))/m}$. DKW inequality only achieves $m^{-1/2}$ rate without the standard deviation factor. 

To capture the scaling of the standard deviation of the $\sqrt{m}\wh{F}_m(w)$, \cite{anderson1952asymptotic} introduced a standardized version  and \cite{eicker1979asymptotic} introduced a studentized version of the Kolmogorov-Smirnov statistic:
$$ T^{\mathrm{AD}}_m := \sup_{w:{F}_W(w)\in(0, 1)} \frac{\sqrt{m}|\wh{F}_m(w) - {F}_W(w)|}{\sqrt{F_W(w)(1-{F}_W(w))}} , \quad T^{\mathrm{Eicker}}_m := \sup_{w:\wh{F}_m(w)\in(0, 1)} \frac{\sqrt{m}|\wh{F}_m(w) - {F}_W(w)|}{\sqrt{\wh{F}_m(w)(1-\wh{F}_m(w))}}.$$
These test statistics are weighted versions of the Kolmogorov-Smirnov statistic, which can be inverted to get confidence bands for $F_W(w)$. These normalized versions are based on the asymptotic normality of $\widehat{F}_m(w)$.  
It should, however, be noted that when $F_W(w)=\ord(1/m)$, $m\wh{F}_m(w)$ has a non-degenerate Poisson distribution limit \citep{le1960approximation, prokhorov1953asymptotic}, suggesting that the width for small (or large) values of $F_W(w)$ can be asymptotically smaller than $\ord(m^{-1/2})$. Confidence bands with such improvements were introduced in \cite{owen1995nonparametric} by inverting the goodness-of-fit tests in \cite{berk1979goodness}. The Berk-Jones statistic is $T^{\mathrm{BJO}}_m := m \sup_{w\in \re}K(\wh{F}_m(w), F_W(w))$ where $K(\cdot, \cdot)$ is the KL divergence between Bernoulli random variables:
$K(a, b) = a\log\left({a}/{b}\right) + (1 - a)\log\left({(1 - a)}/{(1 - b)}\right).$
Let $\kappa_{n,\del}^{\text{BJO}}$ be the $(1-\delta)$-th quantile of $T_m^{\mathrm{BJO}}$, i.e., $\mathbb{P}(T_m^{\mathrm{BJO}} \le \kappa_{n,\del}^{\text{BJO}}) \ge 1 - \delta$. The inequality $m\sup_{w}K(\wh{F}_m(w), F_W(w)) \le \kappa_{n,\delta}^{\text{BJO}}$ results in a confidence band $[\ell_{n,\del}^{\text{BJO}}(w), u_{n,\del}^{\text{BJO}}(w)], w\in\mathbb{R}$. 
\cite{jager2007goodness} proved that the width of this confidence interval asymptotically scales like $2\sqrt{2\gamma_m F_W(w)(1 - F_W(w))}+ 2\gamma_m$ where $\gamma_m = \ord(\log \log m/m)
$. The width rates achieved are $\ord\left(\log\log m/m\right)$ for $F_W(w) = \ord(1/m)$ but give a rate of $\ord\left(m^{-1/2}\sqrt{\log\log m}\right)$ for $F_W(w) = \Omega(1)$, which is slower than the usual rate achieved by DKW inequality. \cite{dumbgen2023new} introduced a modified version of the Berk-Jones statistic by adding an additive correction term $C_\nu(v,t)$, giving the final test statistic: $m \sup_{w\in \re}(K(\wh{F}_m(w), F_W(w)) - C_\nu(\wh{F}_m(w), F_W(w)))$. \cite{dumbgen2023new} prove that the resulting confidence intervals achieve optimal rate. 

\citet[Section 5]{berk1979goodness} showed that the KL divergence-based tests are more powerful than any normalized Kolmogorov-Smirnov statistic in terms of Bahadur slopes. \cite{dumbgen2023new} improve on the Berk-Jones confidence bands by constructing rate-optimal simultaneous confidence intervals. \citet[Section 3]{li2020essential} provide further optimality properties of the confidence sets constructed using the centered KL divergence.

\begin{figure}[H]
\centering
    \includegraphics[width=0.7\textwidth]{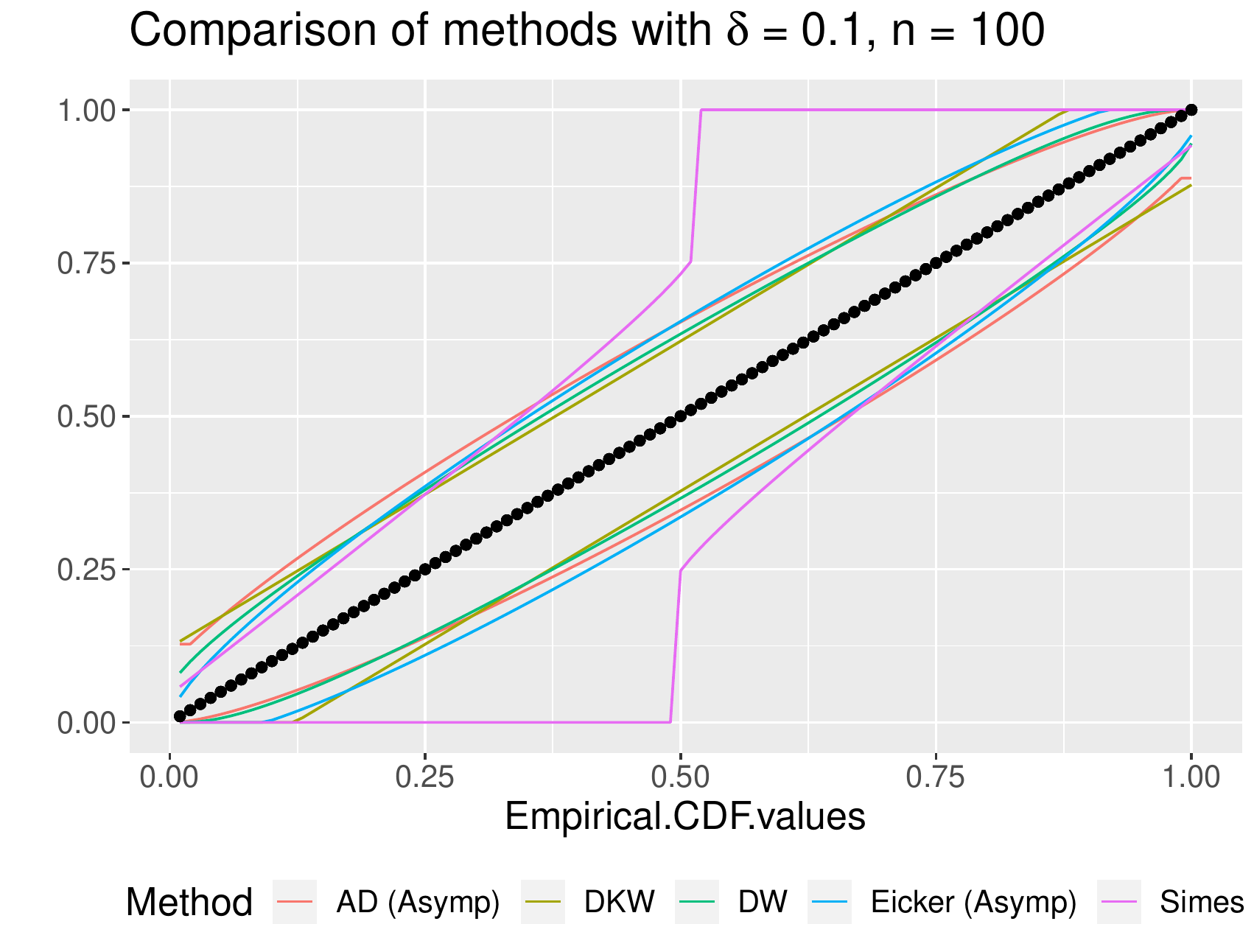}
    \caption{Comparing confidence bands obtained by different methods. The black dots in the center are $\{i/n:\,i\in[n]\}$. For Anderson-Darling and Eicker statistics, we used asymptotic quantiles instead of using Monte Carlo simulations. For the DW method, we set $\nu$ to $3/2$.}
    \label{fig:compare_conf_bands}
\end{figure}

\citet[Proposition 2]{bates2023testing} proposes the use of generalized Simes inequality \citep{sarkar2008generalizing} to construct a confidence band, which tends to be much wider than the traditional confidence bands discussed above. \citet[Section 3.4]{bates2023testing} also considers using the asymptotic distribution of $T_m^{\mathrm{AD}}$. A quick comparison of all these confidence bands is given in Fig~\ref{fig:compare_conf_bands} for $m = 100$ standard uniform random variables.

In the following, we will discuss the DKW and D{\"u}mbgen-Wellner confidence bands. The former is the easiest to use/compute but weaker, while the latter is slightly complicated but sharper. Throughout the discussion below, we fix the training data and the non-conformity score; all the results mentioned below are valid no matter what non-conformity score is used. Further, we use the notation $\wh{F}(t) = |\mathcal{I}_{\mathrm{cal}}|^{-1}\sum_{i\in\mathcal{I}_{\mathrm{cal}}} \mathbf{1}\{s(Z_i) \le t\}.$

\subsection{DKW inequality method}
Recall that the DKW confidence band is $[\ell_{n,\delta}^{\mathrm{DKW}}(t), u_{n,\delta}^{\mathrm{DKW}}(t)]$ with $\ell_{n,\delta}^{\mathrm{DKW}}(t) = \wh{F}(t) - \sqrt{\ln(2/\delta)/(2|\mathcal{I}_{\mathrm{cal}}|)}$ and $u_{n,\delta}^{\mathrm{DKW}}(t) = \wh{F}(t) + \sqrt{\ln(2/\delta)/(2|\mathcal{I}_{\mathrm{cal}}|)}$. By Corollary 1 of~\cite{massart1990tight}, this confidence band satisfies the hypothesis of Theorem~\ref{thm:si-PAC-guarantee} and hence, via Algorithm~\ref{algo:simul-pac} yields a simultaneous $\delta$-PAC prediction sets. Set $\wh{C}_{n,\alpha,\delta}^{\mathrm{DKW}} = \{z:\,s(z) \le \wh{Q}_{n,\alpha,\delta}^{\mathrm{DKW}}\},$ where $\wh{Q}_{n,\alpha,\delta}^{\mathrm{DKW}}$ is the smallest $t$ that satisfies $\ell_{n,\delta}^{\mathrm{DKW}}(t) \ge 1 - \alpha.$

The DKW simultaneous conformal prediction set satisfies a simultaneous coverage guarantee with a probability of at least $1-\delta$. The simultaneity allows the analyst to choose any $\alpha\in(0, 1)$ without losing validity. It would be useful to analyze the probability with which $\cov_P(\wh{C}^{\text{DKW}}_{n,\al,\del})\geq 1 -\al$ \textit{for a fixed $\al$}, or in other words, what is the $\kappa$ for which $\widehat{C}^{\text{DKW}}_{n,\al,\delta}$ is $(\alpha,\kappa)$-PAC valid? 
The following result first provides a bound on $R_{n,\alpha}$ in Theorem~\ref{thm:si-PAC-guarantee} for DKW confidence band and then shows that one can take 
\begin{equation}\label{eq:kappa_DKW_ineq}
\kappa_{\alpha,\delta}^{\text{DKW}} ~:=~ \begin{cases}
(\delta/2)^{1/(4\alpha(1 - \alpha))}, &\mbox{if }\alpha \le 0.5,\\
\delta/2, &\mbox{if }0.5 < \alpha \le 0.5 + \Delta_n^{\text{DKW}},\\
(\delta/2)^{1/(4(\alpha - \Delta_n^{\text{DKW}})(1 - \alpha + \Delta_n^{\text{DKW}}))}, &\mbox{if }\alpha > 0.5 + \Delta_n^{\text{DKW}},
\end{cases}
\end{equation}
where $\Delta_n^{\text{DKW}} := \sqrt{2\ln(2 /\del)/{(9|\mathcal{I}_{\mathrm{cal}}|})} + 1/(3|\Ical|)$. 

\begin{theorem}[Upper bound on coverage and marginal coverage]\label{thm:dkw-upperbd+margcov}
If the non-conformity scores $S_i, i\in\mathcal{I}_{\mathrm{cal}}$ are almost surely distinct, then $R_{n,\alpha} \le \sqrt{2\ln(2/\delta)/|\mathcal{I}_{\mathrm{cal}}|} + 1/|\mathcal{I}_{\mathrm{cal}}|$.
Further, for any $\alpha, \delta \in (0,1)$ and any $n\ge1$, if the distribution of the conformal scores is continuous, then
\begin{align}\label{eq:marg-dkw}
  \Pr\left( \cov_P(\wh{C}^{\mathrm{DKW}}_{n,\al,\del}) \geq 1-\al \right)
~\geq~  1 - \kappa^{\mathrm{DKW}}_{\alpha,\delta} ~\ge~ 1 - \delta/2.
\end{align}

\end{theorem}

The probability bound suggests that the marginal coverage probability is greater than $1-\al$ is the highest when $\al$ is close to $0$ or $1$ and the lowest for $\al = 0.5$; see Fig~\ref{fig:marg_cov}.

\subsection{D\"{u}mbgen-Wellner method}\label{subsec:methods_dw_ineq}

\cite{dumbgen2023new} provides a generalization of the law of iterated logarithm, specifically related to the KL divergence between empirical CDF and true CDF; recall the definition of KL divergence from Section~\ref{subsec:conf_band_litrev}. Unlike~\cite{berk1979goodness} and~\cite{owen1995nonparametric}, this generalized confidence band is constructed using adjusted KL divergence. 
Formally, with non-conformity scores $S_i = s(Z_i), i\in\mathcal{I}_{\mathrm{cal}}$ the test statistic of~\cite{dumbgen2023new} is defined with a (tuning) parameter $\nu > 3/4$ as
\begin{align}\label{eq:DW-test-statistic}
T_{n,\nu}^{\mathrm{DW}} := &
\sup_{
z \in \re}\left\{ |\mathcal{I}_{\mathrm{cal}}|\cdot K( \wh{F}(z), F(z))
- C_\nu( \wh{F}(z), F(z))\right\},
\end{align}
where $C_{\nu}(u,v) := \min\{C(t) + \nu D(t):\,\min(u,v) \leq t \leq \max(u,v)\}$ with
\begin{align*}
&C(t) := \log\left(\log\left(\frac{e}{4t(1 -t)}\right)
\right)\geq  0, \quad
D(t) := \log(1 + C(t)^2) \geq  0.
\end{align*}
Theorem 2.1 of~\cite{dumbgen2023new} shows that $T_{n,\nu}^{\mathrm{DW}}$ has a non-degenerate limiting distribution for any $\nu > 3/4$, when $F(\cdot)$ is continuous. Furthermore, similar to DKW statistic, $T_{n,\nu}^{\mathrm{DW}}$ is also distribution-free for all continuous distributions for the non-conformity scores. For non-continuous distributions, the test statistic, although not distribution-free, is stochastically bounded 
by the continuous version distribution-free statistic (proof in Appendix Section~\ref{pf:stoch_dom_cont_discont}). Thus, we can use Monte-Carlo simulations with standard uniform random variables to obtain accurate estimates of the quantiles. Let $\kappa_n^{\mathrm{DW}}(\del)$ be the $(1-\delta)$-th quantile of $T_{n,\nu}^{\mathrm{DW}}$ computed using Monte-Carlo simulations using the representation of $T_{n,\nu}^{\mathrm{DW}}$ given in Section 3.1 of~\cite{dumbgen2023new}. This yields the confidence band for all $x\in\mathbb{R}$ as $[\ell_{n,\delta}^{\mathrm{DW}}(x),\, u_{n,\delta}^{\mathrm{DW}}(x)]$, which can be written in terms of the order statistics of non-conformity scores as follows. Let $S_1' \leq S_2' \leq \cdots \leq S_{|\mathcal{I}_{\mathrm{cal}}|}'$ represent the order statistics of $S_i, i\in\mathcal{I}_{\mathrm{cal}}$. Also, set $S_0' = -\infty$ and $S_{|\mathcal{I}_{\mathrm{cal}}|+1}' = \infty$. For $0\leq i\leq |\mathcal{I}_{\mathrm{cal}}|$, and $S_{i}' \leq x < S_{i+1}'$, the DW band at $x$ is  
$\ell_{n,\delta}^{\mathrm{DW}}(x) = \ell_{i,n,\del},\, u_{n,\delta}^{\mathrm{DW}}(x) = u_{i,n,\del},$
where  $\ell_{0,n,\del} := 0, u_{|\mathcal{I}_{\mathrm{cal}}|,n,\del} := 1$ and for $0 \leq j < |\mathcal{I}_{\mathrm{cal}}|$, $\ell_{|\mathcal{I}_{\mathrm{cal}}|-j,n,\del} := 1 - u_{j,n,\del}$ with
\begin{equation}\label{eq:bounds_dw} 
\begin{split}
   u_{j,n,\del} &:= \max\left\{u\in(\wh{F}(S_j'), 1]:\, |\mathcal{I}_{\mathrm{cal}}|\cdot K(\wh{F}(S_j'),u) -
C_\nu(\wh{F}(S_j'), u) \leq \kappa_n^{\mathrm{DW}}(\del)\right\}.
\end{split}
\end{equation}
In comparison to the DKW confidence band, the DW confidence band can be much smaller at both endpoints of the distribution. Let $w_{j,n,\del} =  u_{j,n,\delta} - \ell_{j,n,\delta}$ represent the width of the band. \citet[Thms. 3.5-3.7]{dumbgen2023new} shows that $w_{j,n,\delta} \lesssim |\mathcal{I}_{\mathrm{cal}}|^{-1/2}$ for all $j$, similar to the DKW confidence band and similar to~\cite{berk1979goodness} confidence band, $w_{j,n,\delta}\lesssim (\log\log |\mathcal{I}_{\mathrm{cal}}|)/|\mathcal{I}_{\mathrm{cal}}|$ if $\min\{j, |\mathcal{I}_{\mathrm{cal}}|-j\} \lesssim (\log\log |\mathcal{I}_{\mathrm{cal}}|)$. 

By definition, $[\ell_{n,\delta}^{\mathrm{DW}}(x),\,u_{n,\delta}^{\mathrm{DW}}(x)], x\in\mathbb{R}$ is a finite-sample valid confidence band for $F(\cdot)$, and hence, Theorem~\ref{thm:si-PAC-guarantee} implies that $\wh{C}_{n,\alpha,\delta}^{\mathrm{DW}} = \{z:\,s(z) \le \wh{Q}_{n,\alpha,\delta}^{\mathrm{DW}}\},$ where $\wh{Q}_{n,\alpha,\delta}^{\mathrm{DW}}$ is the smallest $t$ that satisfies $\ell^{\mathrm{DW}}_{n,\delta}(t) \ge 1 - \alpha$ is a collection of simultaneous $\delta$-PAC prediction sets. The following result first provides an upper bound on $R_{n,\alpha}$ appearing in Theorem~\ref{thm:si-PAC-guarantee} and then analyzes the probability with which $\cov_P(\widehat{C}^{\mathrm{DW}}_{n,\alpha,\delta}) \ge 1 - \alpha$, or in other words, what is the $\kappa$ for which $\widehat{C}^{\mathrm{DW}}_{n,\al,\delta}$ is $(\alpha,\kappa)$-PAC valid? 
The following result shows that one can take 
\begin{equation}\label{eq:kappa_DW_ineq}
\kappa_{\alpha,\delta}^{\mathrm{DW}} ~:=~ \begin{cases}
    {\exp\left( -C^u_\nu\left(1-\al\right) -\kappa_n^{\mathrm{DW}}\right)}, & \al < 1/2,\\[2ex]
    {\exp\left( -C^u_\nu\left(1-\al+ \Delta_n^{\mathrm{DW}}\right) -\kappa_n^{\mathrm{DW}}\right),} & \al >1/2 + \Delta_n^{\mathrm{DW}},\\[2ex]
    \exp\left(-\kappa_n^{\mathrm{DW}}\right), & \al \in [1/2, 1/2 + \Delta_n^{\mathrm{DW}}],
\end{cases}
\end{equation}
where $C_\nu^u(t) = C(t) + \nu D(t) $ and $\Delta_n^{\mathrm{DW}} = \max_{j \in [|\Ical|]}( (j+1)/|\Ical|  - \ell_{j,n,\del}) $. 

\begin{theorem}[Upper bound on coverage and marginal coverage]\label{thm:dw-upperbd+margcov}
Let  $j_\al = \min\{j\in\{1, 2, \ldots, |\mathcal{I}_{\mathrm{cal}}|\}: \ell_{j,n,\delta}  \geq 1-\al\}$.
If the non-conformity scores $\{S_i\}_{i \in \mathcal{I}_{\mathrm{cal}}}$ are almost surely distinct, then $R_{n,\alpha} \le  w_{j_\al,n,\del} + w_{j_\al-1,n,\del} + 1/|\mathcal{I}_{\mathrm{cal}}|$. Furthermore, for any $\alpha, \delta \in (0,1)$ and any $n\ge1$, if the distribution of the conformal scores is continuous, then
\begin{align}\label{eq:marg-dw}
  \Pr\left( \cov_P(\wh{C}^{\mathrm{DW}}_{n,\al,\del}) \geq 1-\al \right)
~\geq~  1 - \kappa^{\mathrm{DW}}_{\alpha,\delta} ~\ge~ 1 - \exp\left(-\kappa_n^{\mathrm{DW}}\right). 
\end{align}
\end{theorem}

\begin{figure}[h]
\centering
    \includegraphics[width=0.7\textwidth]{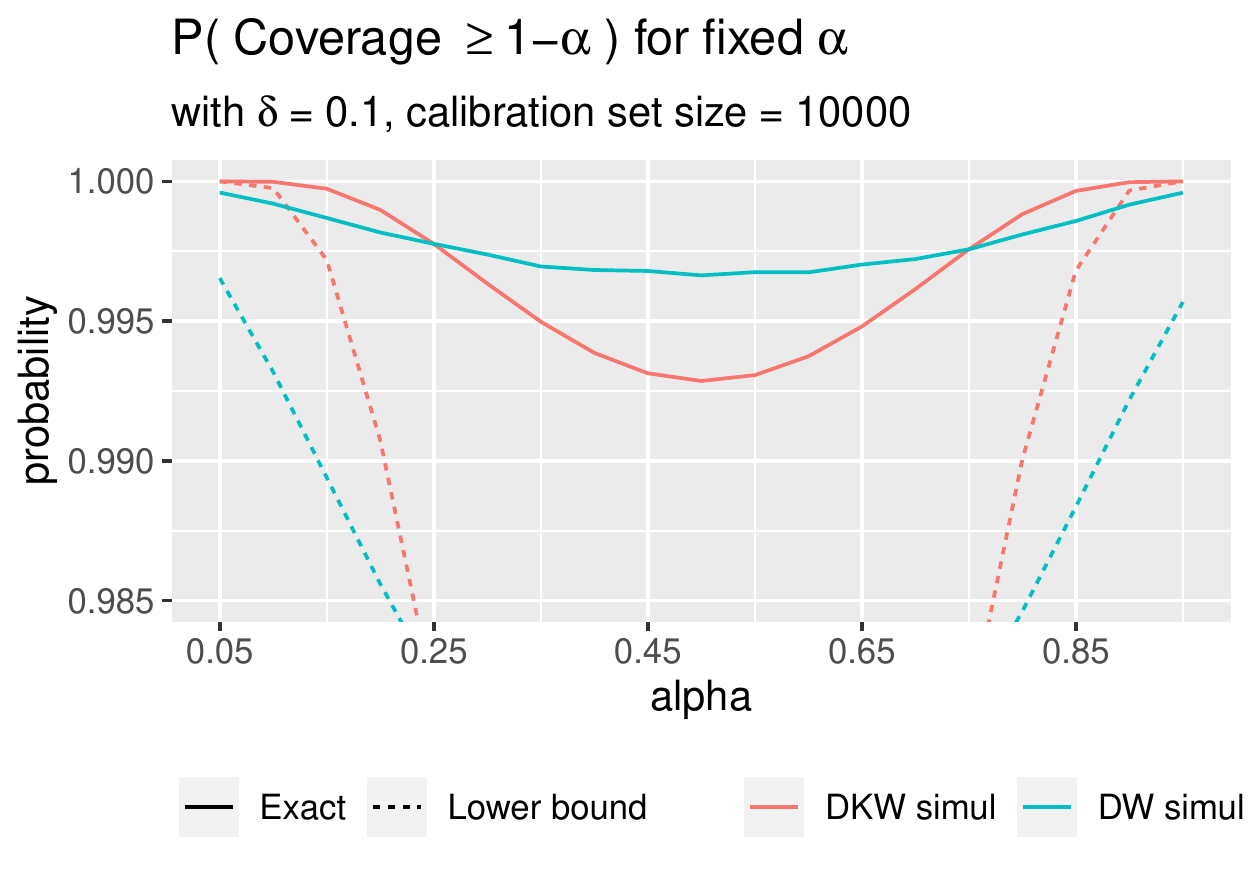}
   \caption{ Analyzing $\prob( \cov(C_\al) \geq 1-\al)$ for a fixed $\al$ for DKW and DW simultaneous prediction sets. We calculate $\prob( \cov(C_\al) \geq 1-\al)$ exactly (shown as solid lines) using the beta distribution of order statistics, and Theorems~\ref{thm:dkw-upperbd+margcov} and \ref{thm:dw-upperbd+margcov} imply that these exact values are lower bounded by $1-\kappa_{\al,\del}$'s (shown as dotted lines). For the DW method, we set $\nu$ to $3/2$.}
 \label{fig:marg_cov}
 \end{figure}

We compare the $\kappa^{\mathrm{DW}}_{\alpha,\delta}$ and $\kappa^{\text{DKW}}_{\alpha,\delta}$ in Fig~\ref{fig:marg_cov}. If $\kappa_{\alpha,\delta}$ from any method is close to $\delta$ (or $\del/2$), then it implies that $([0, 1], \delta)$-PAC simultaneous prediction set is very close to the $(\alpha, \delta)$-PAC prediction set, which in turn implies that there is not much price to pay for simultaneity over $\alpha$. For $\al$ close to $0$ or $1$, $\kappa^{\text{DKW}}_{\alpha,\delta}<\kappa^{\mathrm{DW}}_{\alpha,\delta}$ and furthermore, 
$\kappa^{\text{DKW}}_{\alpha,\delta}$ approach $0$ at a faster rate than $\kappa^{\mathrm{DW}}_{\alpha,\delta}$ in these regions. This shows that the D\"{u}mbgen-Wellner method would be more favorable in the usual settings where the coverage level is high ($\al$ is low), as $\kappa_{\al,\del}$ would be closer to $\delta$. \par 

\paragraph{Rivera-Walther Confidence set.}The confidence bands discussed in Section~\ref{subsec:conf_band_litrev} are based on the empirical and the true cumulative distribution function. The adjusted KL divergence approach of~\cite{dumbgen2023new} can also be applied to construct an optimal confidence set for the probability measure (not just the CDF). This, in turn, leads to a confidence band. We now provide a brief discussion of this optimal confidence set based on Section 3.1 of \cite{li2020essential}. The adjusted KL divergence is now computed between $\widehat{F}_n(I)$ and $F(I)$ for some finite intervals $I$ instead of using $I = (-\infty, x]$. We call this the Rivera-Walther confidence set because the sequence of intervals $I$ used was introduced in~\cite{rivera2013optimal}. This sequence is not only sparse ($\ord(n)$ number of intervals) but also sufficiently rich to capture information. For notational convenience, we describe this method based on IID observations $X_1, \ldots, X_n$ from the distribution $F$ and we use the notation $F(I) = \mathbb{P}(X\in I)$. Define $l_{\max} = \lfloor \log_2(n/\log n)\rfloor$ and for $2 \le l\le l_{\max}$, set $m_{l} = n2^{-l}$ and $d_l = \lceil m_l/(6l^{1/2})\rceil$. Let $X_{(1)} \le X_{(2)} \le \ldots \le X_{(n)}$ denote the order statistics. With $\mathcal{D} = \{i:\, X_{(i)} \neq X_{(i+1)}\}$, define
\begin{align*}
\mathcal{J} =  \bigcup_{l=2}^{l_{\max} } \mathcal{J}(l),\quad\mbox{where}\quad
\mathcal{J}(l) =  \left\{(X_{(j)}, X_{(k)}]: j,k \in \{1 + id_l, i \in \N_0\}\cap \mathcal{D},\quad m_l<k-j< 2m_l \right\}.
\end{align*}
The test statistic used for the confidence set is  
 \begin{align*} \label{eq:essHist}
T_n^{\text{RW}} =  \max_{I \in \mathcal{J}} \ \left\{\sqrt{2n K\big(F(I), F_n(I)\big)}- \sqrt{\mathfrak{C}(F_n(I))}\right\},
\text{ where } 
\mathfrak{C}(t) =  2 \log\left(\frac{e}
{t(1 - t)}
\right).
\end{align*}
Note that $\mathfrak{C}(\cdot)$ is almost the same as $C(\cdot)$ used in~\eqref{eq:DW-test-statistic}.
Let $\kappa_n^{\text{RW}}(\del)$ be the quantile of $T_n^{\text{RW}}$. Now, using this distribution, the $(1-\delta)$-confidence set for the probability measure is given by
$$   \left\{H\mbox{ a probability measure}:\, \sqrt{2nK\big(H(I), F_n(I)\big)}- \sqrt{\mathfrak{C}(F_n(I))}
\leq \kappa_n^{\text{RW}}(\del) \quad \forall I \in \mathcal{J} \right\}.$$
Several optimality properties of this confidence set are proved in Section 4 of~\cite{li2020essential}.
Inverting the KL divergence, this confidence set can be rewritten as
$$ C_n^{\text{RW}}(\del) = \{ H\text{ a
probability measure}: \ell_{n,\delta}(I) \leq H(I) \leq u_{n,\delta}(I) \quad \forall I \in \mathcal{J} \}.$$ 
Note that since $I$ is of the form $(X_{(j)}, X_{(k)}]$, the inequalities are of the form $$ \ell_{kj} \le F(X_{(k)}) - F(X_{(j)}) \le u_{kj} \quad \forall k>j \ s.t.\ (X_{(j)}, X_{(k)}] \in \mathcal{J}$$ 
These bounds on difference could be used to obtain bounds on $\{ F(X_{(i))}\}_{i \in [n]}$ This can be formulated as a linear program for finding lower bounds (or upper bounds) for each $i$.  
\begin{equation*}
\begin{array}{ll@{}ll}
\text{minimize }  &L^n_i & \\
\text{subject to}& \ell_{kj} \le L^n_k - L^n_j \le u_{kj} & \quad  \forall k>j \ s.t.\ [X_{(j)}, X_{(k)}) \in \mathcal{J}\\
& 0 \leq L^n_j \leq 1 , &  \quad j=1 ,\dots, n
\end{array}
\end{equation*}
This LP essentially looks at the lowest possible value of $F(X_{(i)})$ allowed by the inequalities in the Rivera-Walther confidence set. 
\begin{figure}[H]
\centering
\includegraphics[width=10cm]{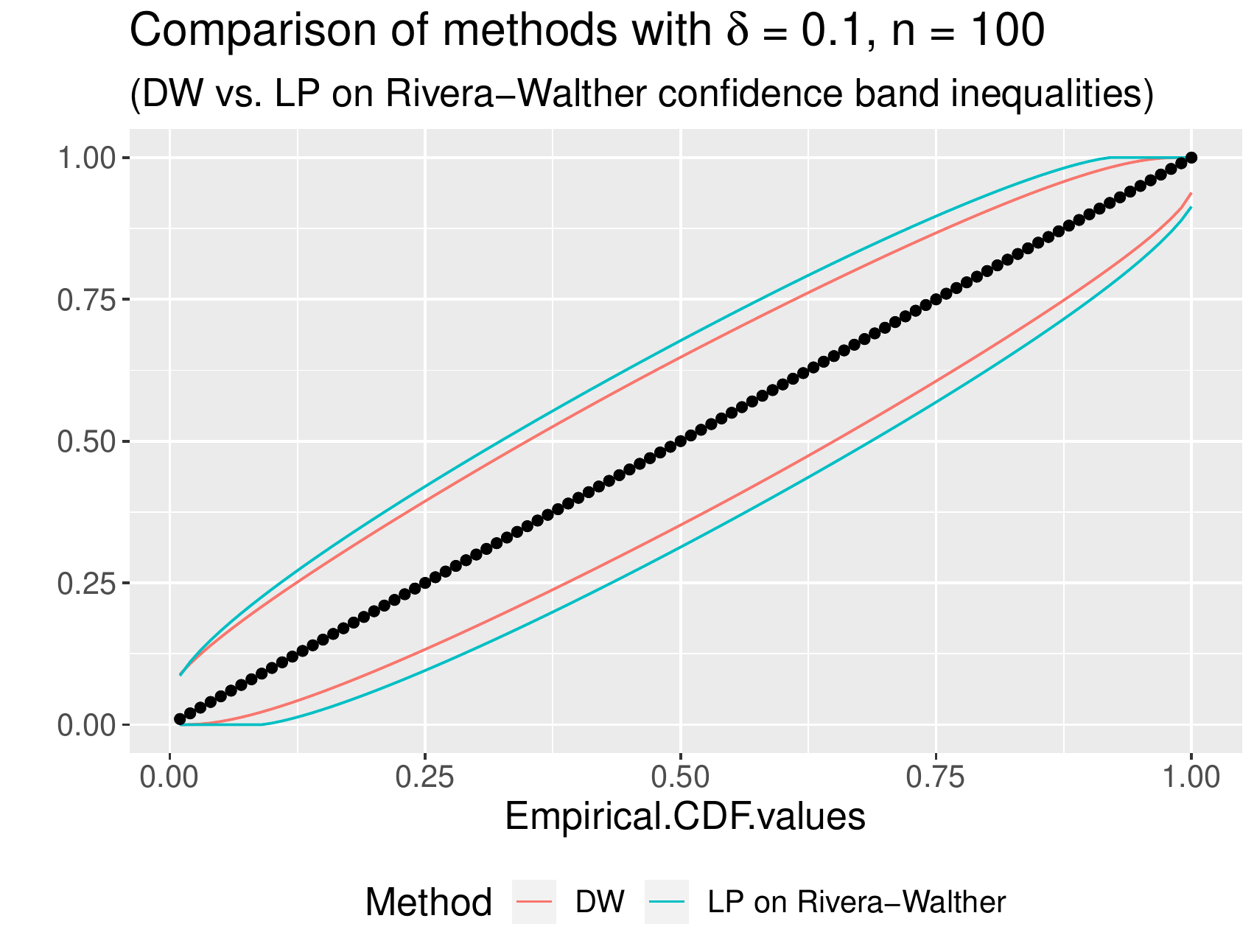}
\caption{Comparing confidence bands obtained by \cite{dumbgen2023new} and \cite{rivera2013optimal}. The black dots in the center are $\{i/n:\,i\in[n]\}$. For both methods, we set $\nu$ to $3/2$.} 
\label{fig:compare_rivera_walther_dw}
\end{figure}

We compare the bounds obtained by applying the LP above on an essential histogram statistic on all possible intervals of the form $(X_{(j)},X_{(k)}]$, instead of the sparse set of intervals \citep{rivera2013optimal} in Fig~\ref{fig:compare_rivera_walther_dw}; the test statistic with all possible intervals is denoted by $P_n^{all}$ in~\cite{rivera2013optimal}. Both confidence bands follow a similar shape, but the confidence bounds obtained via the Rivera-Walther confidence set approach are wider. This is to be expected since \cite{li2020essential} provides an optimal confidence band for $F$ on all sets of the form $\left(X_{(j)},X_{(k)}\right]$, whereas \cite{dumbgen2023new} only provides confidence bands on sets of the form $\left(-\infty,X_{(k)}\right]$ i.e. $F\left(X_{(k)}\right)$. Therefore, the essential histogram approach will be more conservative in providing a confidence band for $F\left(X_{(k)}\right)$.

\section{Simulated and Real data Applications}\label{sec:simulations-real-data}
We now apply the methods discussed in two settings: when the response is continuous (regression) and categorical (classification). We compare the following conformal prediction methods
\begin{enumerate}[leftmargin=*]
    \item \textbf{Training data quantile method:} Here we compute the non-conformity scores on the training data itself and compute the $(1-\alpha)$ quantile of them. This method will not have a joint coverage guarantee, since the non-conformity scoring method is dependent on the training dataset. 
   \item \textbf{Split conformal:} Here we split the data into training data and calibration data.  The non-conformity scoring method is made using quantile regressors (or classifiers) on the training dataset. And the scores are computed for the calibration set. We use the conformal score from \cite{chernozhukov2021distributional} for regression (see Example~\ref{eg:dcp}) and from~\cite{romano2020classification} for classification. These methods have joint coverage guarantees but no simultaneous $\delta$-PAC guarantee.   
    \item \textbf{Simultaneous PAC methods:} We apply Algorithm~\ref{algo:simul-pac} using the DKW and DW confidence bands on the same non-conformity scoring methods as mentioned above. Instead of the simultaneous $\delta$-PAC guarantee, we choose ($I = [0,0.5]$, $\delta$)-PAC i.e. the conformal sets provided would be valid simultaneous for all $\al\in[0.0.5]$. This is not a limitation because in practice one only uses prediction sets with coverage between $0.5$ and $1$. For the DW method, $\nu$ is set to $3/2$.
\end{enumerate}
We compare the methods using two metrics:
\begin{enumerate}[leftmargin=*]
   \item \textbf{Coverage:} We consider how the coverage \textit{varies across} $\al$ for each of these methods. For clarity, we plot $\cov_P(\widehat{C}_{n,\alpha}) - (1 - \alpha)$ as a function of $\alpha$ for different prediction sets $\widehat{C}_{n,\alpha}$ with $\cov_P(\widehat{C}_{n,\alpha})$ approximated using test data as ${ \sum_{i \in \mathcal{I}_{\text{test}}} \one( Y_i \in  \wh{C}_\al(X_i))}/{| \mathcal{I}_{\text{test}}|}$.
   \item \textbf{Precision:} We also consider the precision (width) of the methods across $\al$: $\text{Size}(\wh{C}_\al) \approx { \sum_{i \in \mathcal{I}_{\text{test}}} \text{Size}(\wh{C}_\al(X_i))}/{| \mathcal{I}_{\text{test}}|}$. For regression, $\text{Size}$ refers to the length of the prediction interval. For classification, $\text{Size}$ refers to the number of elements in the prediction set. 
\end{enumerate}

\subsection{Simulated data}\label{appsec:sim_data}
\subsubsection{Dataset and method}
For simulated data, we look at the perturbed Poisson distribution model used in \cite{romano2019conformalized}. The covariates are $X_i\sim \text{Unif}\ [1,5]$ and the response $Y_i$ is defined as follows.
$$Y_i = \text{Pois}(\sin^2
(X_i) + 0.1) + 0.03 X_i + \eps_{1,i} + \mathbbm{1} (U_i < 0.01)\eps_{2,i} $$
where $\eps_{1,i}, \eps_{2,i}$ are IID standard normal random variables and $U_i\sim \text{Unif} [0,1]$. We opt for the DCP split conformal method \citep{chernozhukov2021distributional} in this setting. This method involves estimating $\prob( Y\leq y| X = x )$ for which we use quantile regression random forests \citep{meinshausen2006quantile}. 
\par

The goal is to demonstrate the PAC guarantees of the proposed methods compared to the standard methods. To achieve this, we run 100 simulations with the output of each simulation iteration giving us a random $\wh{C}_\al$ for $\al$ ranging from 0.05 to 0.95.
 Within each simulation iteration, we generate a new training set (of size $50000$) and a new calibration set (of size $10000$) and run the entire procedure to obtain a prediction set $\wh{C}_\al$. 

\subsubsection{Coverage}
\begin{figure}[h]
    \includegraphics[width=\textwidth]{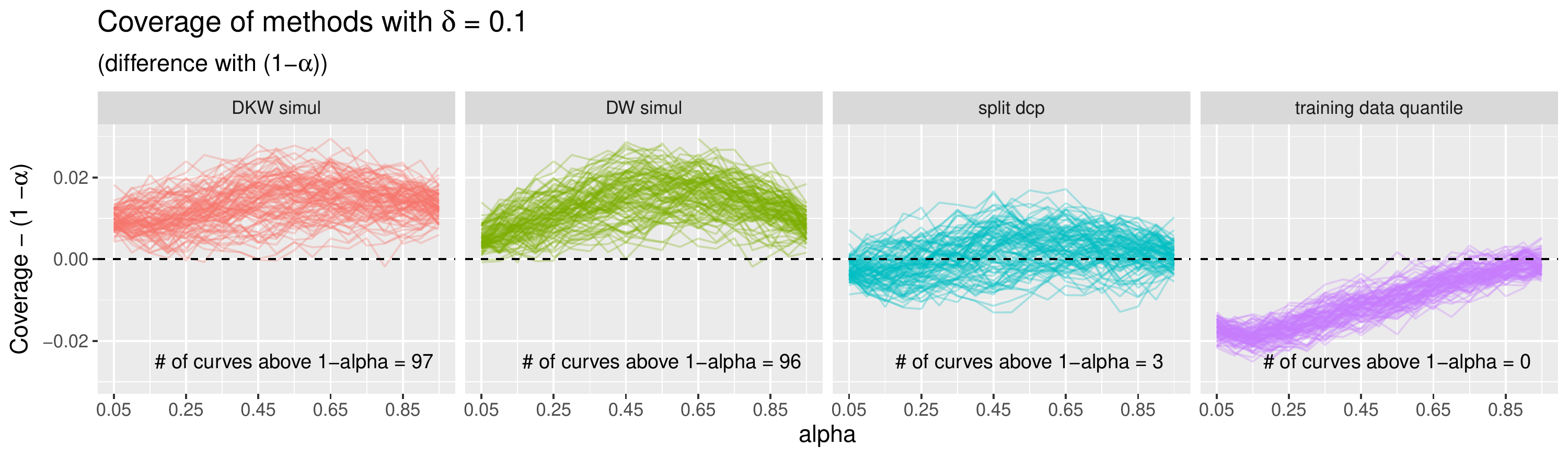}
    \caption{Coverage across different methods. Each curve is coverage probability for different $\al$, for each simulation (total $100$). The number at the bottom is an estimate of $\prob(\cov_P(\wh{C}_\al) \geq 1-\al 
    \ \forall \al )$. Methods \textit{DKW simul} and \textit{DW simul} offer simultaneous PAC guarantees and the number should be close to $100(1-\del)$.  For the DW method, we set $\nu$ to $3/2$.}  
    \label{fig:sim_cov}
\end{figure}

For each simulation, we approximate $\cov_P(\wh{C}_\al)$ by looking at empirical coverage over a test dataset of size $20000$ i.e. 
 $\cov_P(\wh{C}_\al) \approx { \sum_{i \in \mathcal{I}_{\text{test}}} \one( Y_i \in  \wh{C}_\al(X_i))}/{| \mathcal{I}_{\text{test}}|}$.
As shown in Fig~\ref{fig:sim_cov}, the coverage probability in the case of simultaneous PAC guaranteed DCP is above $1-\al$ around $100(1- \del)$ times as expected. The standard split conformal DCP coverage is symmetric around $1-\al$, demonstrating that it achieves joint coverage. And as expected, the training data quantile method underperforms.\par

\subsubsection{Width}

\begin{figure}[H]
\includegraphics[width=\textwidth]{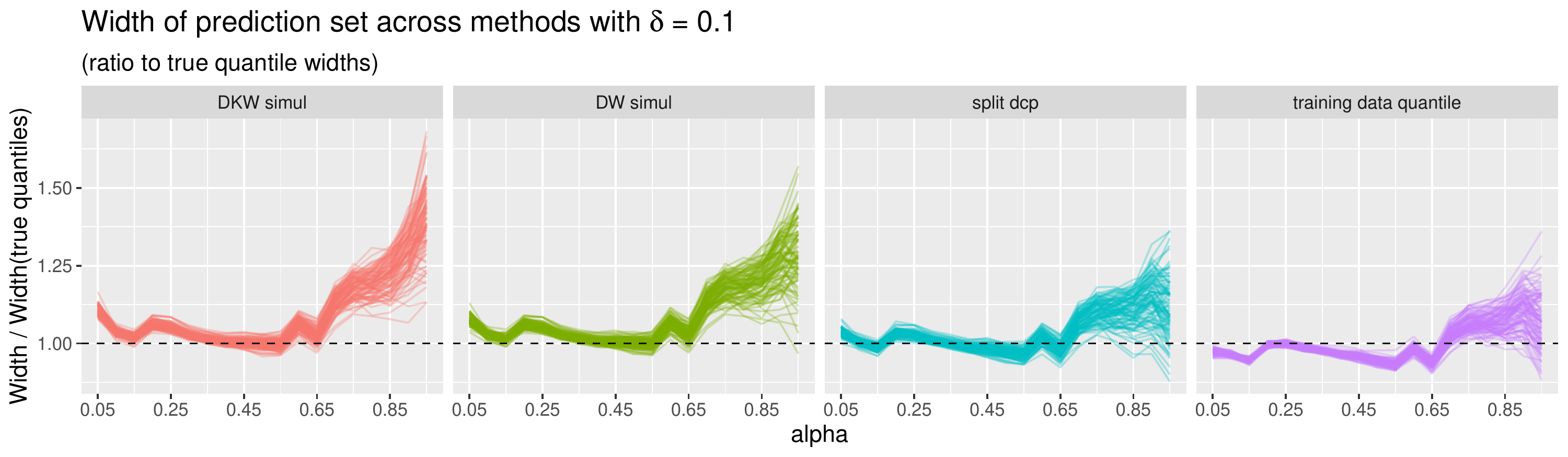}    
\caption{Width of prediction set across different methods. Each curve is the ratio of the average width for each method to true quantiles on the test dataset for different $\al$ and for each simulation (total $100$).  For the DW method, we set $\nu$ to $3/2$.} 
\label{fig:sim_width}
\end{figure}

Although there is no closed form for the quantiles of $Y|X$, the data generation process is known. Hence, one can estimate them well via Monte Carlo estimation. We use the width of these quantiles as a benchmark to compare with the widths of different methods. For each simulation, we compute the average width of the prediction set on the test dataset i.e.
 $\text{Width}(\wh{C}_\al) \approx  \sum_{i \in \mathcal{I}_{\text{test}}} \text{Width}(\wh{C}_\al(X_i))/{|\mathcal{I}_{\text{test}}|}$.
In Fig~\ref{fig:sim_width}, we see that the width of the prediction sets from the simultaneous PAC methods is mostly wider than the true quantile width (ratio $> 1$) for each alpha. This is in accordance with a PAC guarantee requirement. The standard split conformal DCP has a width ratio hovering around $1$, in line with its joint coverage guarantee. And the training data quantiles have a shorter width, reflective of the bias due to obtaining quantiles  of non-conformity scores on the training set itself. Also, we know that the DW method provides confidence bands that are narrower than the DKW inequality method for extreme values of the CDF. This is demonstrated in Fig~\ref{fig:sim_width}, where the DW inequality method provides narrower width of prediction sets for $\al \in\{ 0.05,0.95\}$.

\subsection{Classification: Fashion MNIST}\label{subsec:fashionMNIST}
\subsubsection{Dataset and method.} 
We use the Fashion MNIST dataset, introduced in \cite{xiao2017fashion}. The dataset is a collection of gray-scale images of fashion items of 10 categories. Each image is $28\times28$ pixels, which we convert into a vector of length 784. The whole dataset is of size 70000, which we split randomly into training (size $50000$), calibration (size $10000$), and test (size $10000$) datasets. We opt for the classification split conformal method introduced in \cite{romano2020classification} in this setting (Example~\ref{eg:class}). This method involves an estimator of $\Pr(Y = y| X = x)$, for which we use random forests \citep{breiman2001random}. We know that \cite{romano2020classification} involves picking labels with the highest estimated conditional probability $Y|X$. Thus, for a low enough coverage probability ($\al$ large), the conformal prediction set will only have a single element which is the highest estimated probability label.
\subsubsection{Coverage and width.}
\begin{figure}[h]
    \centering
    \includegraphics[width=\textwidth]{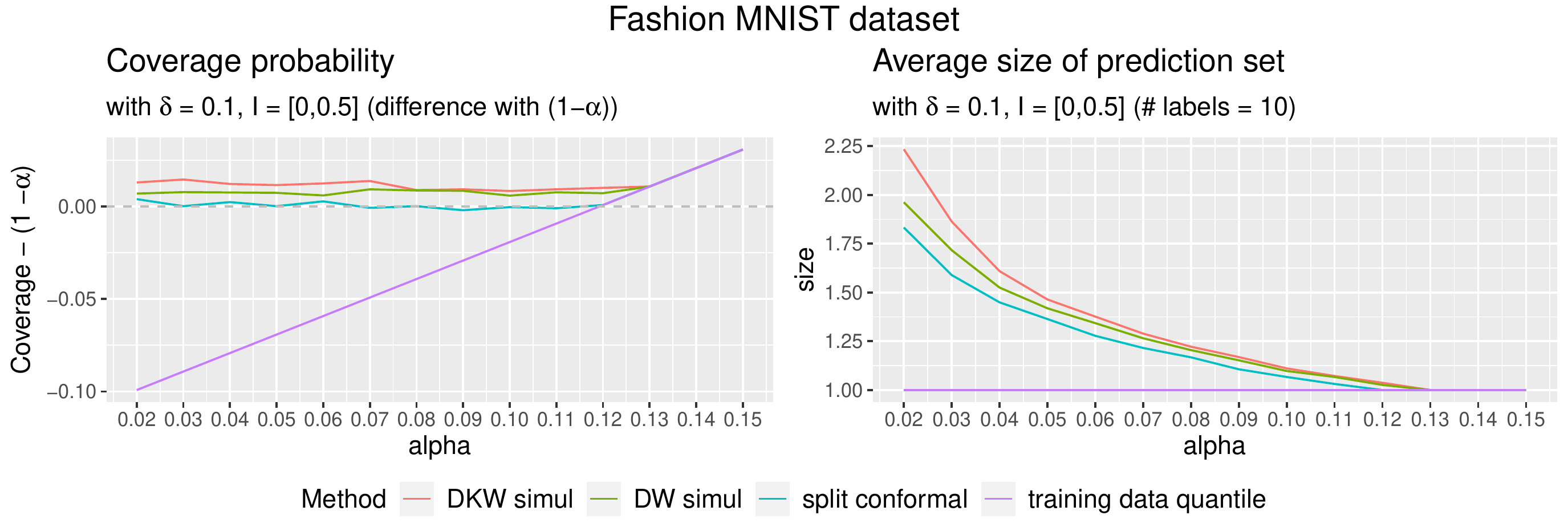}\\    
    \caption{Coverage (left) and average size (right) across different methods for each $\al$ for the Fashion MNIST dataset. We restrict the plot to $\al\le 0.15$ for a clearer comparison. For the DW method, we set $\nu$ to $3/2$.}
    \label{fig:real_class}
\end{figure}

As discussed above, for large values of $\al$, the $\wh{C}_\al$ will be the highest probability label, which is the prediction of the classifier. Thus, for large $\al$ the $\cov_P(\wh{C}_\al)$ will be equal to the test accuracy of the classifier, and the size of the prediction set is also 1. We observe this in Fig~\ref{fig:real_class}, where for $\al \geq 0.13$, all the conformal prediction sets have coverage equal to the test accuracy of $0.8808$. 
The simultaneous methods have coverage greater than $1-\al$, while the split and training quantile methods fail to satisfy the simultaneous guarantee. The DW simultaneous PAC sets dominate the DKW simultaneous PAC sets both in terms of having coverage closer to the nominal $1-\al$ and in terms of having a smaller \text{Size}.  

    \label{fig:real_class_size}

\subsection{Regression: Airfoil dataset}\label{subsec:real_data_reg}
\subsubsection{Dataset and method.}
We use the Airfoil dataset, obtained from NASA \citep{brooks1989airfoil}. The dataset contains results of aerodynamic and acoustic tests conducted on airfoil blade sections under different controlled experimental conditions. Each datapoint has covariates related to the experiment conditions, with the response variable being the scaled sound pressure. We split the dataset randomly into a training dataset of size $700$, a calibration dataset of size $300$, and a test dataset of size $503$. We opt for the distributional split conformal method introduced in \cite{chernozhukov2021distributional} in this setting (Example~\ref{eg:dcp}). This method involves estimating $\prob( Y\leq y| X = x )$ for which we use random forests \citep{meinshausen2006quantile}. 
\subsubsection{Coverage and width.}
The conclusions are similar to those observed in the Fashion MNIST dataset. Additionally, here we note that the DW simultaneous PAC prediction sets have better (closer to nominal) coverage and smaller width for $\alpha \le 0.25$, while the DKW simultaneous PAC prediction sets dominate for $\alpha \ge 0.25$. This is expected from Figure~\ref{fig:marg_cov}.

\begin{figure}[H]
    \centering
    \includegraphics[width = \textwidth]{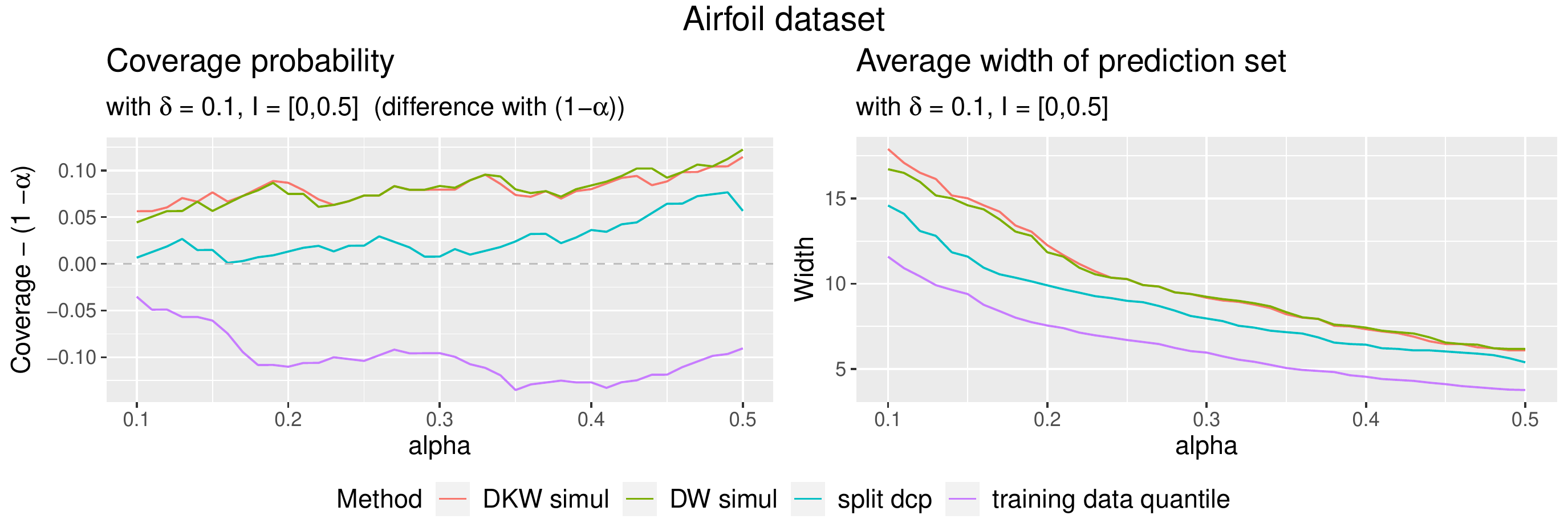}\\
    \caption{Coverage (left) and average width (right) across different methods for each $\al$ for the airfoil dataset. For the DW method, we set $\nu$ to $3/2$.}
    \label{fig:real_reg}
\end{figure}
\section{Conclusion}\label{sec:conclusion}
In this paper, we formulated the problem of constructing prediction sets with nominal coverage that is data-dependent. This is a specific instance of (predictive) inference-after-selection. The flexibility of choosing a data-dependent coverage level of a prediction set yields practitioners to trade-off precision (width/size of prediction sets) and accuracy (coverage of prediction sets). This is, especially, important in the context of classification problems where a high coverage requirement might yield trivial prediction sets and the analyst wishes to reduce the coverage requirement. We have proved the equivalence of this problem to the problem of constructing confidence bands for a cumulative distribution function. We introduce two different simultaneous PAC prediction sets based on DKW inequality and optimal confidence bands of~\cite{dumbgen2023new}. The latter confidence band is known to be rate-optimal, and hence, is expected to yield the ``optimal'' simultaneous coverage guarantees; Note that there is no single optimal confidence band because lowering the width at one point often results in an increase of width at another point. 

In this paper, we have restricted ourselves to independent and identically distributed (IID) data, while most literature on conformal prediction only requires exchangeable data. It is unclear to us if the methods in this paper extend readily to exchangeable data. Another interesting direction is to extend the methods to covariate shift data, which was embedded into a missing data framework with IID observations in~\cite{yang2022doubly}. We believe that the methods described extend to this setting. 
\bibliographystyle{apalike}
\bibliography{ref}

\newpage
\appendix

\section[]{Stochastic dominance of $T_{n,\nu}^{\mathrm{DW}}$ (continuous $F$ over discrete $F$)}\label{appsec:stochastic-dominance}

\begin{proof}\label{pf:stoch_dom_cont_discont}
 Let $G$ be the distribution of a discrete distribution, with points of discontinuities at $u_1<u_2 \ldots < u_m $ with probabilities  are $p_1,p_2, \ldots p_m$ respectively. There exists a continuous CDF $F$ such that $F(u_i) - F(u_{i-1}) = p_i \ \forall i \in [m]$ with $u_0 = -\infty$. By construction, note that $F(u_i) = G(u_i) \ \forall i \in [m]$. Therefore, for any random variable $X \sim F$, $Y = \sum_{i= 1}^m u_i \one(X \in(u_{i-1},u_i]) \sim G $. Given this relationship, any sample $\{Y_i\}_{i \in [n]}\sim G$ can be seen as coming from $\{X_i\}_{i \in [n]}\sim F$. Since $Y_i \in \{u_i \}_{i \in [m]}$ we can say for any $i \in [m]$
 $$ \sum_{j=1}^n I( Y_j \leq u_i ) = \sum_{i=j}^n I( X_j \leq u_i ) \implies 
 \wh{G}_n(u_i) = \wh{F}_n(u_i). $$
 This implies that
 \begin{align*}
T_{\nu}^{\mathrm{DW}}(G) &= 
\sup_{
u_i \in [m] }\left\{ nK( \wh{G}(u_i), G(u_i))
- C_\nu(\wh{G}(u_i), G(u_i)) \right\}\\
&=\sup_{
u_i \in [m] }\left\{ nK( \wh{F}(u_i), F(u_i))
- C_\nu(\wh{F}(u_i), F(u_i)) \right\}\\
&\leq \sup_{
u \in \re }\left\{ nK( \wh{F}(u), F(u))
- C_\nu(\wh{F}(u), F(u)) \right\} = T_{\nu}^{\mathrm{DW}}(F).\\
 \end{align*}
 Note that $T_{\nu}^{\mathrm{DW}}(F)$ is distribution-free, proving the result.
\end{proof}

\section{Proofs}
\subsection{Proof of Theorem~\ref{thm:si-PAC-guarantee}} 
\begin{proof}

    Let $F$ be the CDF of $s(Z)$.  By the provided confidence band, we know that 
    
    $$\prob\left(\ell_{n,\delta}(t) \le F(t) \le u_{n,\delta}(t)\;\forall\;t\in\mathbb{R}\right) \ge 1 - \delta $$
We substitute $t \in \re$ with $\{ \widehat{Q}_{n,\alpha,\delta}^{\mathrm{sim}}: \al \in (0,1)\}$. We can say that  
 $$\prob\left(\ell_{n,\delta}(\widehat{Q}_{n,\alpha,\delta}^{\mathrm{sim}}) \le F(\widehat{Q}_{n,\alpha,\delta}^{\mathrm{sim}}) \le u_{n,\delta}(\widehat{Q}_{n,\alpha,\delta}^{\mathrm{sim}})\;\forall\; \al \in (0,1)\right) \ge 1 - \delta.$$
 Since $\cov_P( \wh{C}^{\text{sim}}_{n,\al,\del}) = F( \widehat{Q}_{n,\alpha,\delta}^{\mathrm{sim}} )$ and $ 1- \al \leq  \ell_{n,\del}(\widehat{Q}_{n,\alpha,\delta}^{\mathrm{sim}})$, we can instead write
$$\prob\left( 1-\al \le\cov_P( \wh{C}^{\text{sim}}_{n,\al,\del})\le u_{n,\delta}(\widehat{Q}_{n,\alpha,\delta}^{\mathrm{sim}})\;\forall\; \al \in (0,1)\right) \ge 1 - \delta.$$
By definition of $R_{n,\al}$, we get that $u_{n,\delta}(\widehat{Q}_{n,\alpha,\delta}^{\mathrm{sim}}) = 1 - \alpha + R_{n,\alpha}$ and therefore 
$$
\mathbb{P}\left(1 - \alpha \le \cov_P(\widehat{C}_{n,\alpha}^{\mathrm{sim}}) \le 1 - \alpha + R_{n,\alpha}\;\forall\; \al \in (0,1)\right) \ge 1 - \delta.$$
\end{proof}


\subsection{Proof of Theorem~\ref{thm:dkw-upperbd+margcov}} 
\begin{proof}

Let $F$ be the CDF of $s(Z)$, then $\cov_P( \wh{C}^{\text{DKW}}_{n,\al,\del}) = F( \widehat{Q}_{n,\alpha,\delta}^{\mathrm{DKW}} )$. For $S_i = s(Z_i)$ for $i \in \mathcal{I}_{\text{cal}}$. For simplicity, let $m = |\mathcal{I}_{\text{cal}}|$ and let $S_1' \le S_2' \le \cdots \le S_m'$ represent the (increasing) order statistics of $\{S_i\}, i\in\mathcal{I}_{\mathrm{cal}}$ (note that these values are almost surely distinct when $F$ is continuous). Let $\eps^m_{\del} = \sqrt{\ln(2 /\del)/{2m}}  $, then $(\ell_{n,\delta}^{\mathrm{DKW}}(S'_j),u_{n,\delta}^{\mathrm{DKW}}(S'_j)) = (\widehat{F}(S'_j) -\eps^m_{\del},\widehat{F}(S'_j) + \eps^m_{\del})$  and therefore
$$\widehat{Q}_{n,\alpha,\delta}^{\mathrm{DKW}} = S_{j_\al^m}' \text{ where } j_\al^m = \min\left\{j: \wh{F}(S_j')\geq 1 - \alpha + \eps^m_{\del}\right\} .$$
Since $j_\al^m$ is defined as a minimum, we conclude 
\begin{align*}
\label{eq:j_al_dkw}
1 - \alpha + \eps^m_{\del} & \leq  \wh{F}(S_{j_\al^m}')\ ,\  1 - \alpha + \eps^m_{\del}  > \wh{F}(S_{j_\al^m - 1}')\\
\implies \widehat{F}(S'_{j_{\alpha}^m}) &= \widehat{F}(S'_{j_{\alpha}^m - 1}) + \frac{1}{m}\sum_{j=1}^m \mathbf{1}\{S'_{j} = S'_{j_{\alpha}^m}\}\\ 
&\le 1 - \alpha + \eps^m_{\del} + \frac{1}{m}\sum_{j=1}^m \mathbf{1}\{S'_{j} = S'_{j_{\alpha}^m}\}.
\end{align*}
If $S_i'$ are distinct then $\sum_{j=1}^m \mathbf{1}\{S'_{j} = S'_{j_{\alpha}^m}\} = 1$, thus giving us $ \widehat{F}(S'_{j_{\alpha}^m})  \le 1 - \alpha + \eps^m_{\del} + 1/m$. Analyzing $R_{n,\al}$, we get that
 \begin{align}
 R_{n,\alpha} & = [\ell_{n,\delta}^{\mathrm{DKW}}(S_{j_\al^m}') - (1 - \alpha)] + [u_{n,\delta}^{\mathrm{DKW}}(S_{j_\al^m}') - \ell_{n,\delta}^{\mathrm{DKW}}(S_{j_\al^m}')]\\
 & = [\widehat{F}(S'_{j_{\alpha}^m}) - \eps^m_{\del} - (1-\al)] + [2\eps^m_\del] \leq \frac{1}{m} + 2\eps^m_\del .
 \end{align}
Given $\widehat{Q}_{n,\alpha,\delta}^{\mathrm{DKW}} = S_{j_\al^m}'$, we can write the following equivalence

$$ \Pr\left( \cov_P\left(\wh{C}^{\text{DKW}}_{n,\al,\del}\right) \geq 1- \al\right)  =  \Pr\left( F\left(S_{j_\al^m}'\right) \geq 1- \al\right). $$

Because the distribution of the non-conformity scores is continuous, we can say that $\wh{F}(S_j') = j/m$ almost surely and $F(S_{(j_{\alpha}^m)}')$ is identically distributed as $j_{\alpha}^m$-th order statistic of $m$ IID standard uniform random variables. This implies that $F(S_{j_\al^m}')\sim
\text{Beta}(j_\al^m, m - j_\al^m + 1)$. Corollary 8 of \cite{henzi2023some}, for  any $X\sim \text{Beta}(a,b)$ and $x \in [0,  p_l]$ where $p_l= a/(a+b-1)$
\begin{equation}\label{eq:corollary-8-Henzi-Dumbgen}
\prob( X\leq x ) \leq \exp\left(
-\frac{(a + b -1)(x - p_l)^2}{2(2x/3 + p_l/3)(1- 2x/3 -p_l/3)}\right),
\end{equation}
Taking $a = j_{\alpha}^m$ and $b = m - j_{\alpha}^m + 1$, we get $p_l = j_\al/m$. We take $x = 1-\al$.
By choice of $j_\al^m$, we know that $ 1 - \alpha  + \eps_\del^m \leq  j_\al^m/m \leq 1-\al +  \eps_\del^m + 1/m$, thus satisfying the condition required for~\eqref{eq:corollary-8-Henzi-Dumbgen}. 
Thus, 
\begin{align*}
 \Pr\left( F\left(S_{j_\al^m}'\right) \leq 1- \al\right) & \leq \exp\left(
-\frac{m(1-\al -j_\al^m/m  )^2}{2(2(1-\al)/3 + j_\al^m/(3m) ))(1-2(1-\al)/3 -j_\al^m/(3m) )}\right)   
\end{align*}
We now look at the term inside the exponential. Let $\Delta_n^{\text{DKW}} = \eps_\del^m/3 + 1/(3m)  $
\begin{itemize}
    \item \textbf{Denominator:}
$2(1-\al)/3 + j_\al^m/(3m) \in [1-\al, 1-\al + \Delta_n^{\text{DKW}} ] $, and we can say that 
\begin{align*}
&2(2(1-\al)/3 + j_\al^m/3m  ))(1-2(1-\al)/3 -j_\al^m/3m )\\ 
&\qquad\leq \begin{cases}
    2\al(1-\al) & \al < 1/2\\
    2(\al - \Delta_n^{\text{DKW}} )(1-\al + \Delta_n^{\text{DKW}})     & \al > 1/2 + \Delta_n^{\text{DKW}}\\
    1/2 & \al \in [1/2, 1/2 + \Delta_n^{\text{DKW}}]
\end{cases}
\end{align*}
\item \textbf{Numerator:} $m(1-\al -j_\al^m/m  )^2 \geq m(\eps_\del^m)^2 = \ln(2/\del)/2$
\end{itemize}
This gives the final bound 
$$ \Pr\left( F\left(S_{j_\al^m}'\right) \leq 1- \al\right)
\leq  \begin{cases}
     (\del/2)^{(4\al(1-\al))^{-1}} & \al < 1/2\\
     (\del/2)^{(4(\al-\Delta_n^{\text{DKW}})(1-\al + \Delta_n^{\text{DKW}}))^{-1}}    & \al >1/2 + \Delta_n^{\text{DKW}}\\
    \del/2 & \al \in [1/2, 1/2 + \Delta_n^{\text{DKW}}]
\end{cases}$$
Noting $\mathbb{P}\left(\cov_P(\widehat{C}_{n,\alpha,\delta}^{\text{DKW}}) \ge 1 - \alpha\right) = 1 - \mathbb{P}\left(F(S_{j_\al^m}') < 1 - \alpha\right)$, we obtain the result.

\end{proof}


\subsection{Proof of Theorem~\ref{thm:dw-upperbd+margcov}}
\begin{proof}
Let $F$ be the CDF of $s(Z)$, then $\cov_P( \wh{C}^{\mathrm{DW}}_{n,\al,\del}) = F( \widehat{Q}_{n,\alpha,\delta}^{\mathrm{DW}} )$. For $S_i = s(Z_i)$ for $i \in \mathcal{I}_{\text{cal}}$. For simplicity, let $m = |\mathcal{I}_{\text{cal}}|$ and let $S_1' \le S_2' \le \cdots \le S_m'$ represent the (increasing) order statistics of $\{S_i\}, i\in\mathcal{I}_{\mathrm{cal}}$ (note that these values are almost surely distinct when $F$ is continuous).
$$\widehat{Q}_{n,\alpha,\delta}^{\mathrm{DW}} = S_{j_\al^m}' \text{ where } j_\al^m = \min\left\{j: \ell_{j,n,\delta}\geq 1 - \alpha\right\}  $$
Using the almost sure distinctness of the non-conformity scores, we can say that $\wh{F}(S_j') = j/m$ almost surely. By construction of the bands, $\wh{F}(S_j') = j/m \in [ \ell_{j,n,\delta}, u_{j,n,\delta}]$, giving us that 
\begin{align}\label{eq:DW-bound-l_j}
    &u_{j-1,n,\delta} \geq  (j- 1)/m \ , \ \ell_{j,n,\delta} \leq j/m\\
 \implies &    \ell_{j,n,\delta} \leq u_{j-1,n,\delta}+ 1/m\\
\end{align}

As defined before, let $w_{j,n,\del} =  u_{j,n,\delta} - \ell_{j,n,\delta}$. Since $j_\al^m$ is defined as a minimum, we conclude that $
1 - \alpha  \leq \ell_{j_\al^m,n,\delta}$ and $ \ell_{j_\al^m-1,n,\delta}  <  
1 - \alpha $, this gives us
\begin{align}
 R_{n,\alpha} & = [\ell_{n,\delta}^{\mathrm{DW}}(S_{j_\al^m}') - (1 - \alpha)] + [u_{n,\delta}^{\mathrm{DW}}(S_{j_\al^m}') - \ell_{n,\delta}^{\mathrm{DW}}(S_{j_\al^m}')]\\
 & = [ \ell_{j_\al^m,n,\del} - (1-\al)] +  w_{j_\al^m,n,\del} \\
 & \leq  [\ell_{j_\al^m,n,\del} - \ell_{j_\al^m-1,n,\del}] +  w_{j_\al^m,n,\del} \\
& \leq  [u_{j_\al^m-1,n,\del} + 1/m - \ell_{j_\al^m-1,n,\del}] +  w_{j_\al^m,n,\del} \quad \hbox{(By Eq~\eqref{eq:DW-bound-l_j})}\\
 & \leq  w_{j_\al^m,n,\del}  + w_{j_\al^m - 1,n,\del} + 1/m.
 \end{align}
 
Given $\widehat{Q}_{n,\alpha,\delta}^{\mathrm{DW}} = S_{j_\al^m}'$, we can write the following equivalence.

$$ \Pr\left( \cov_P\left(\wh{C}^{\mathrm{DW}}_{n,\al,\del}\right) \geq 1- \al\right)  =  \Pr\left( F\left(S_{j_\al^m}'\right) \geq 1- \al\right)  $$

As in the proof of Theorem~\ref{thm:dkw-upperbd+margcov}, we know $F(S_{j_\al^m}')\sim
\text{Beta}(j_\al^m, m - j_\al^m + 1)$. Theorem 5 of \cite{henzi2023some}, for  any $X\sim \text{Beta}(a,b)$ and $x \in [0,  p_l]$ where $p_l= a/(a+b-1)$
\begin{equation}\label{eq:thm-5-Henzi-Dumbgen}
\prob( X\leq x )  \leq \exp\left(
-(a+b-1)K(p_l,x)\right),
\end{equation}
Taking $a = j_{\alpha}^m$ and $b = m - j_{\alpha}^m + 1$, we get $p_l = j_\al/m$. We take $x = 1-\al$.
By choice of $j_\al^m$, we know that $ 1 - \alpha  \leq  j_\al^m/m $, thus satisfying the condition required for~\eqref{eq:thm-5-Henzi-Dumbgen}. We use the  definition of $j_\al$ as follows,
\begin{align*}
&1-\al \leq L_{j_\al,\del}^m\\
\implies & mK\left(\frac{j_\al}{m}, 1-\al\right) -  C_\nu\left(\frac{j_\al}{m}, 1 - \al \right)
> \kappa_n^{\mathrm{DW}}(\del).
\end{align*}
Note that $C_{\nu}(u,v) := \min\{C^u_\nu(t):\min(u,v) \leq t \leq \max(u,v)\}   $. Given $j_\al/m \geq 1- \al$, we can say that 
$$ C_\nu\left(\frac{j_\al}{m}, 1 - \al \right) = \begin{cases}
  C_\nu^u\left(1-\al\right) 
 & 1 - \al > 1/2\\
 C_\nu^u\left({j_\al}/{m}  \right)
 & {j_\al}/{m} < 1/2  \\
  0 & \hbox{otherwise}
   \end{cases}
$$

We know that ${j_\al}/{m}\in[1-\al, 1 - \alpha + j_\al^m/m - \ell_{n,\delta}^{\mathrm{DW}}(S_{j_\al^m -1}')  ]$. Let $\Delta_n^{\mathrm{DW}} = \max_{j \in [m]} \left( (j+1)/m -\ell_{j,n,\del}\right)$, if ${j_\al}/{m} \leq 1 - \al  +\Delta_n^{\mathrm{DW}}< 1/2 \implies  C_\nu^u\left({j_\al}/{m}  \right) \geq  C_\nu^u\left( 1-\al+\Delta_n^{\mathrm{DW}}\right) $
Thus, we get the following lower bound.

$$ mK\left(\frac{j_\al}{m}, 1-\al\right) >  \begin{cases}
  C_\nu^u\left(1-\al\right) 
 + \kappa_n^{\mathrm{DW}}(\del)& \al<1/2\\
 C_\nu^u\left( 1-\al+\Delta_n^{\mathrm{DW}}\right)
  + \kappa_n^{\mathrm{DW}}(\del)& \al > 1/2 + \Delta_n^{\mathrm{DW}} \\
   \kappa_n^{\mathrm{DW}}(\del)& \al \in[1/2 , 1/2+ \Delta_n^{\mathrm{DW}}] 
   \end{cases}
$$
This gives us the final bound.
$$ \Pr\left( F\left(S_{j_\al^m}'\right) \leq 1- \al\right)
\leq  \begin{cases}
    {\exp\left( -C_\nu^u\left(1-\al\right) -\kappa_n^{\mathrm{DW}}\right)} & \al < 1/2\\[2ex]
    {\exp\left( -C_\nu^u\left(1-\al+ \Delta_n^{\mathrm{DW}}\right) -\kappa_n^{\mathrm{DW}}\right)} & \al >1/2 + \Delta_n^{\mathrm{DW}}\\[2ex]
    \exp\left(-\kappa_n^{\mathrm{DW}}\right) & \al \in [1/2, 1/2 + \Delta_n^{\mathrm{DW}}]
\end{cases}$$
Noting $\mathbb{P}\left(\cov_P(\widehat{C}_{n,\alpha,\delta}^{\text{DW}}) \ge 1 - \alpha\right) = 1 - \mathbb{P}\left(F(S_{j_\al^m}') < 1 - \alpha\right)$, we obtain the result.
\end{proof}
\end{document}